\RequirePackage[hyphens]{url}
\documentclass[letterpaper,twocolumn,10pt]{article}
\usepackage{usenix}
\usepackage{hyperref}
\usepackage{amsfonts, amsmath, amssymb, amsthm}
\usepackage{graphicx, tikz} 
\usetikzlibrary{positioning,arrows.meta}

\usepackage{algorithm}
\usepackage[noend]{algpseudocode} 

\theoremstyle{definition}
\newtheorem{definition}{Definition}

\theoremstyle{theorem}
\newtheorem{theorem}{Theorem}

\title{Extending the Formalism and Theoretical Foundations of Cryptography to AI}

\author{
\begin{tabular}[t]{c}
Federico Villa\\
ETH Zurich\\
\texttt{fvilla@student.ethz.ch}
\end{tabular}
\and
\begin{tabular}[t]{c}
F.~Bet\"{u}l Durak\thanks{Corresponding author.}\\
Microsoft Research\\
\texttt{betul.durak@microsoft.com}
\end{tabular}
\and
\begin{tabular}[t]{c}
Tadayoshi Kohno\\
Georgetown University\\
\texttt{yoshi.kohno@georgetown.edu}
\end{tabular}
\and
\begin{tabular}[t]{c}
Tapdig Maharramli\\
ETH Zurich\\
\texttt{tmaharramli@student.ethz.ch}
\end{tabular}
\and
\begin{tabular}[t]{c}
Franziska Roesner\\
University of Washington\\
\texttt{franzi@cs.washington.edu}
\end{tabular}
}

\begin{document}
\newcommand{\betul}[1]{{\color{blue}[Bet\"{u}l: #1]}}
\newcommand{\federico}[1]{{\color{orange}[Federico: #1]}}
\newcommand{\tapdig}[1]{{\color{brown}[Tapdig: #1]}}
\newcommand{\franzi}[1]{{\color{magenta}[Franzi: #1]}}
\newcommand{\yoshi}[1]{{\color{purple}[Yoshi: #1]}}

\newcommand{\algorithmautorefname}{Algorithm}

\newcommand{\finetune}{\textsf{Finetune}}
\newcommand{\reinforce}{\textsf{Reinforce}}
\newcommand{\corpus}{\textsf{corpus}}
\newcommand{\labeleddata}{\textsf{labeledDataset}}
\newcommand{\rankedoutput}{\textsf{rankedOutput}}
\newcommand{\aioracle}{\textsf{AIOracle}}
\newcommand{\vect}{\textsf{vector}}
\newcommand{\LEARN}{\textsf{LEARN}}
\newcommand{\INFER}{\textsf{INFER}}
\newcommand{\result}{\textsf{result}}
\newcommand{\data}{\textsf{corpus}}
\newcommand{\context}{\textsf{context}}
\newcommand{\ATK}{\textsf{ATK}}
\newcommand{\truth}{\textsf{source}}
\newcommand{\query}{\textsf{Q}}
\newcommand{\aux}{\textsf{aux}}
\newcommand{\accept}{\textsf{accept}}
\newcommand{\answer}{\textsf{A}}
\newcommand{\bcoin}{\textsf{b}}      
\newcommand{\bprime}{\textsf{b'}}    
\newcommand{\pipeline}{\textsf{pipeline}}
\newcommand{\state}{\textsf{state}}
\newcommand{\view}{\textsf{view}}
\newcommand{\pvar}{\textsf{p}}
\newcommand{\xzero}{\textsf{context}_0}
\newcommand{\xone}{\textsf{context}_1}
\newcommand{\xb}{\textsf{context}_{\textsf{b}}} 
\newcommand{\y}{\textsf{y}}
\newcommand{\model}{\textsf{model}}
\newcommand{\prompt}{\textsf{prompt}}
\newcommand{\emptyv}{\textsf{empty}} 
\newcommand{\aacm}{\textsf{AACM}}

\newcommand{\seeLearn}{\textsf{see\_learn}}
\newcommand{\injectLearn}{\textsf{inject\_learn}}
\newcommand{\blackbox}{\textsf{black\_box}}
\newcommand{\seePipelineInfer}{\textsf{see\_model}}
\newcommand{\injectInfer}{\textsf{inject\_infer}}

\newcommand{\genEntryForPhase}{\textsf{GENERATE\_DATA}}
\newcommand{\genEntryForPhasei}{\genEntryForPhase\mbox{$_{i}$}}
\newcommand{\learnPhase}{\textsf{learn\_phase}}
\newcommand{\adversary}{\mathcal{A}}
\newcommand{\advantage}{\textsf{Adv}}

\newcommand{\SecurityGame}{\textsf{Security\_Game}}

\maketitle

\begin{abstract}
Recent progress in (Large) Language Models (LMs) has enabled the development of autonomous LM‑based agents capable of executing complex tasks with minimal supervision. 
These agents have started to be integrated into systems with significant autonomy and authority. 
The security community has been studying their security.
One emerging direction to mitigate security risks is to constrain agent behaviours via access control and permissioning mechanisms.
Existing permissioning proposals, however, remain difficult to compare due to the absence of a shared formal foundation.

This work provides such a foundation. 
We first systematize the landscape by constructing an attack taxonomy tailored to language models, the computational primitives of agentic systems. 
We then develop a formal treatment of agentic access control by defining an \aioracle{} algorithmically and introducing a security‑game framework that captures completeness (in the absence of an adversary) and adversarial robustness. 
Our security game unifies confidentiality, integrity, and availability within a single model. 
Using this framework, we show that existing approaches to confidentiality of training data fundamentally conflict with completeness. 
Finally, we formalize a modular decomposition of helpfulness and harmlessness objectives and prove its soundness, in order to enable principled reasoning about the security of agentic system designs.

Our studies suggests that if we were to design a secure system with measurable security, then we might want to use a modular approach to break the problem into sub-problems and let the composition on different modules complete the design.
Our studies show that this natural approach with the relevant formalism is needed to prove security reductions.
\end{abstract}


\section{Introduction}
Recent advances in Artificial Intelligence (AI), including the rapid maturation of (Large) Language Models (LMs), have enabled capabilities which were previously unattainable in deployed systems.
Building on this technological shift, LM-based agents have quickly emerged as a powerful concept: programs that autonomously act on behalf of a human user once given an instruction.
Agents have seen accelerated integration in AI systems due to their abilities to delegate and execute complex tasks with minimal human supervision.
As these systems increasingly act with a high degree of autonomy and authority, the security community has naturally turned its attention to studying their security, for example in relation to the access control permissions granted to such agents \cite{tsai2025contextual, shi2025progent, costa2025securing, camelpaper}.

The agent permissioning space has already seen several similar and intuitively justified proposals. 
However, their constructions can at times feel insufficiently anchored in a common formal foundation.
For example, it is difficult to compare systems like Progent \cite{shi2025progent} and Conseca \cite{tsai2025contextual}: although both aim to regulate agent behaviour through policy mechanisms, Progent allows user-driven policy updates in response to prompts, whereas Conseca adops a more structured dual-agent approach.
Our formalism can explain some of the choices made in these proposals that we analyzed.
Concretely, we clarify the role of post-filtering with our formalism and reason about where the policy is, even when the user adapts the policy, and how it is updated.

\noindent
\textbf{Background, the role and value of formal definitions.}
While we investigate the security foundations of agentic access control mechanisms, our work will adopt an explicitly interdisciplinary perspective with team members spanning AI, systems security, and cryptography. 
In particular, we leverage the rigor of modern cryptography, the formal definitions, adversarial models, and provable security guarantees, to establish a principled basis for evaluating and securing agentic access control architectures built on top of language models.


In early days of modern cryptography, algorithms were proposed based on heuristic arguments and were considered secure until a new attack discovered that they could not withstand.
At the time, there was no rigorous method to precisely measure the security.
Thanks to the solid foundation from complexity theory, we now have formal definitions and reduction-based proofs.
Similarly, to reason about the security of LMs, we need formal definitions that isolate the specific properties these systems must satisfy in order them to be considered secure.

Nowadays, primitives are usually defined by algorithms, a notion of correctness, and a notion of security.
The correctness requirement characterizes the expected behaviour of these algorithm under honest execution in the absence of adversarial interference.
In contrast, the security requirement formulates the unexpected behaviour even when the algorithms are executed in the presence of an adversary employing arbitrary but computationally bounded strategies.
We further define security by a game that an adversary ``plays'' with various capabilities such as access to some information or oracles in a black-box manner.
We quantify the advantage of an adversary and say that a system is secure if every efficient adversary has a negligible advantage.
We can then formally reduce the security of a cryptographic system to a well studied computational problem.
This methodology has a long history in cryptography and forms the foundation of provable security.
With that spirit, we will design a security model in the form of algorithmic games for AI.


AI designs typically specify behavioral objectives in terms of \emph{helpfulness} and \emph{harmlessness}.
These objectives are often intertwined during the the language-model training process, where optimization procedures attempt to improve them simultaneously.
To develop a formal view, we begin by distinguishing two baseline requirements: an AI system should behave as expected in the absence of adversarial influence, and it should not exhibit unexpected behavior when interacting with an adversary constrained by well‑defined capabilities.
We capture these properties using two complementary games: completeness, which formalizes correct behavior under benign conditions, and security, which characterizes robustness against adversarial manipulation.\footnote{We name the notion as completeness, instead of correctness, to avoid confusion with the narrower notion of ``correct output''; completeness reflects the conformity to the system's functional specifications, which is typically more than providing a result which is correct, e.g. ``usefulness''.}

For security, we will define an all-in-one game that unifies confidentiality, integrity, and availability into a single game.
We further observe that the helpfulness and harmlessness objectives can be defined as a conjunction of various criteria that can be achieved separately.
This defines a modular approach which is in line with current practices.
We argue that establishing such modular split is especially valuable when we seek to ``prove'' the security of a design which relies on another design, much as is standard practice in cryptography.

\noindent
\textbf{Our contributions.} We begin with surveying the existing work in agentic systems security, including safety guardrails, model-training interventions, and agentic access control.
Prior approaches treated safety and security separately. 
However, in \autoref{sec:aioracles}, we will treat safety and security together as any problem which would prevent a decrease in both helpfulness and harmlessness. 
In \autoref{sec:taxonomy} and \ref{sec:attack_categories}, we construct an attack taxonomy that systemizes  known classes of attacks against LMs to clarify the landscape.
Later in \autoref{sec:aacm}, we investigate the agentic access control mechanism proposals to demonstrate different approaches.
We, then, transition to a formal treatment in \autoref{sec:formalism}.
We define AI systems algorithmically, articulate a security game framework which captures different levels of security that design choices can provide, along with conditions and assumption required for each.
We prove that current approach to confidentiality of training data is incompatible with utility.
We formalize the modular approach and prove its soundness in terms of completeness and security.

\section{Language Models as \aioracle{s}} \label{sec:aioracles}


In this section, we introduce the \aioracle{} abstraction as a unifying model that captures a broad class of AI systems, including LMs, chatbots, classifiers, and agents. 
This abstraction enables us to reason systematically about their learning and inference behaviours, as well as their safety and security.

\begin{figure}[t]
  \centering
  \resizebox{\columnwidth}{!}{%
  \begin{tikzpicture}[
    node distance=1cm,
    >=Latex,
    stage/.style={
      rectangle, rounded corners, draw, thick,
      minimum height=8mm, minimum width=18mm, align=center
    },
    lab/.style={font=\small}
  ]
    \node[stage] (learn) {\textsf{Learn}};
    \node[stage, right=of learn] (finetune) {\finetune};
    \node[stage, right=of finetune] (reinforce) {\reinforce};
    \node[align=center, right=of reinforce] (model) {\model};
    \node[stage, right=of model] (infer) {\textsf{Infer}};
    \node[right=of infer] (vector) {\vect};

    \draw[->, thick, dashed] (learn) -- (finetune);
    \draw[->, thick, dashed] (finetune) -- (reinforce);
    \draw[->, thick, dashed] (reinforce) -- (model);
    \draw[->, thick, dashed] (model) -- (infer);
    \draw[->, thick, dashed] (infer) -- (vector);

    \node[lab, above=7mm of learn] (lab_corpus) {\data };
    \draw[->] (lab_corpus) -- (learn.north);

    \node[lab, above=7mm of finetune] (lab_labeled) {\labeleddata};
    \draw[->] (lab_labeled) -- (finetune.north);

    \node[lab, above=7mm of reinforce] (lab_ranked) {\rankedoutput};
    \draw[->] (lab_ranked) -- (reinforce.north);

    \node[lab, above=7mm of infer] (lab_input) {\prompt};
    \draw[->] (lab_input) -- (infer.north);
  \end{tikzpicture}%
  }
  \caption{Data and algorithm pipeline in LMs. In \aioracle{}, we will merge \finetune{} and \reinforce{} with \textsf{Learn} into \LEARN{}. Iterated application of \textsf{Infer} will be represented in \INFER{} algorithm where the first output becomes \result{}.}
  \label{fig:pipeline}
\end{figure}
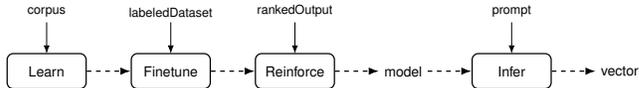

A Language Model (LM) is a set of algorithms which works in two phases: learning and inference. 
The learning phase is a multi-round process.
The algorithms are learning, finetuning and reinforcement learning as we will describe shortly. 
Training process inputs some data called corpus and outputs the first model.
Then the finetuning process takes this model along with some labeled data to output an updated model. 
Finally, reinforcement learning phase further trains the model with some ranked outputs to output the final model.
On the other hand, there is only one algorithm running in inference phase: which inputs a query from a user and uses the model to output a mapping from the dictionary (or vocabulary) to a probability.
Overall, given the token (a word, part of a word, or punctuation) probabilities, an LM algorithms select the next token in the sequence based on this distribution.
While we detailed description of LM algorithms in \autoref{appendix:llm_details} to be complete, 
we draw the pipeline of LM algorithms with their in/outputs in \autoref{fig:pipeline}. 

\subsection{The \aioracle{} Abstraction.} 
In literature, we have different but related notions such as a chatbot, a classifier, and an agent.
A chatbot is based on LMs with iterated inferences.
Instead of only outputting the next token, a chatbot makes complete sentences.
A classifier is very similar to an LM except that the output of the inference algorithm is in a much smaller set of categories.
Finally, an agent can be defined as ``the result of inference procedures outputs a code to be executed in an environment.'' \footnote{We will give a concrete example in \autoref{subsec:aioracle_example}.}
Given these notions, we end up on a notion of an ``\aioracle'' (based on commonality).
Without loss of generality, we say that an \aioracle{} works in two phases: \LEARN{} and \INFER{} which we will clarify and detail later in \autoref{sec:formalism}: \LEARN{} creates a \model{} from a \data{} and \INFER{} computes a \result{} to a \prompt{} using the \model{} (and a \context).

An \aioracle{} is designed with two specific objectives: to be \emph{helpful} and to be \emph{harmless} to the user, society or even to the provider. 
The first difficulty is that helpfulness requires two notions: the result of the query must be \emph{correct}, and the result must be \emph{useful}. If the user asks their very secure \aioracle{} ``what is my next meeting?" and the result is "it is the one which follows your current meeting," it is correct but useless. 
In multi-purpose and complex systems like \aioracle{s}, neither correctness nor usefulness can be algorithmically defined. 
Defining what is useful is specifically harder because we need an end user to judge which might not work well in real-world.
 
\subsection{\aioracle{} Safety \& Security Measurements} 
\begin{figure}[t]
  \centering
  \resizebox{\columnwidth}{!}{%
  \begin{tikzpicture}[
    font=\small\sffamily,
    node distance=8mm,
  ]
    \node at (0, 3.0) {\textbf{Objectives}};
    \node at (4.2, 3.0) {\textbf{Security}};
    \node at (8.4, 3.0) {\textbf{Source}};

    \node (harmlessness) at (0, 1.4) {helpfulness};
    \node (helpfulness)  at (0,0.4) {harmlessness};

    \node (confidentiality) at (4.2,  2.0) {confidentiality};
    \node (integrity)       at (4.2,  1) {integrity};
    \node (availability)    at (4.2, 0.0) {availability};

    \node (training)  at (8.4,  2.0) {training data};
    \node (userdata)  at (8.4,  1.5) {user data};
    \node (truth)     at (8.4, 1) {truth};
    \node (ethics)    at (8.4, 0.4) {ethics};

    \draw (harmlessness) -- (confidentiality);
    \draw (harmlessness) -- (integrity);
    \draw (harmlessness) -- (availability);

    \draw (helpfulness) -- (confidentiality);
    \draw (helpfulness) -- (integrity);
    \draw (helpfulness) -- (availability);

    \draw (confidentiality) -- (training);
    \draw (confidentiality) -- (userdata);

    \draw (integrity) -- (truth);
    \draw (integrity) -- (ethics);

  \end{tikzpicture}%
  }
  \caption{Objectives–Security–Source mapping of \aioracle{}.}
  \label{fig:goals-problems-how}
\end{figure}
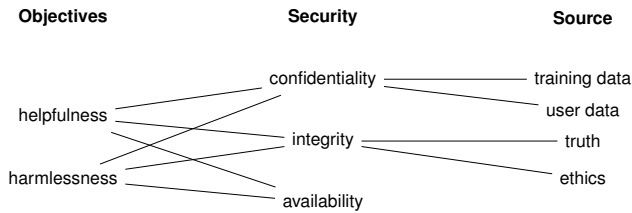

While the helpfulness and harmlessness objectives are intuitive and important to clearly specify, they are not treated independently (for valid reasons), not easily measurable, and highly context-dependent. 
These two objectives are ``tested'' with scientific studies as well as empirical analysis through safety and security measurements.
Before we detail them, we have a layout of objectives and security safety problems as depicted in \autoref{fig:goals-problems-how}.

\noindent
\textbf{Helpfulness and Harmlessness (H\&H).} 
H\&H objectives are treated interdependently.
Defining what H\&H means relates to the difficulty to establish the truth and ethics, which is deeply philosophical. 
Without regulatory oversight, the model developer or provider occupies a unique authoritative position over what constitutes truth and ethical behavior.
To give an example, consider Grok AI which is a chatbot developed by xAI. 
It was designed with ``rebellious streak personality'', humor, and sarcasm much like \emph{The Hitchhiker's Guide to the Galaxy by Douglas Adams}.  
 
H\&H goals mismatch and might go one against the other.
Imagine a user querying ``Which race should I avoid interacting?". 
LMs should refuse to answer with ``You should absolutely avoid Vogons who are the most annoying people.", even if it is an ``accurate'' answer \footnote{This example uses ``Vogons,” a fictional alien species from Adams’ The Hitchhiker’s Guide to the Galaxy, which are portrayed in the novel as notably unpleasant and annoying. The reference is employed solely as a humorous metaphor and not intended to map onto any real‑world human groups. The point is to illustrate that even seemingly ``accurate” answers can still be harmful or inappropriate; it should not be read as involving race or racism.}.
However, it translates into not achieving its helpfulness goal (to an assumed-to-be honest user) as the result will not be what the user wants.
In such conflicting cases, LMs favoring no harm over ``helping an honest user'' is aligned with its design principals and user satisfaction will be irrelevant.
In short, H\&H is defined relative to the design objectives.
Different institutes or governments could come up with their own objectives and principals such as positive bias, censorship, or even an alternative truth.
Simply put, if the models are trained to output biased or incorrect information such as ``elephants can fly,'' the user getting exposed to such information is aligned with the objectives.

The notion we focus as harmlessness here is that broadly, \aioracle{} should not offend, discriminate, or spread hate.
However, it doesn't only limit to non-offensiveness, it also defends against \aioracle{} giving wrong results (such as $13+27 = 3.14$ which would harm a student if they believe it is true) and also maintain privacy to prevent harms that come from leakage.

In our formalism, we will assume that the helpfulness and harmlessness of a result is defined by a predicate $\phi$.
In practice, helpfulness and harmlessness depend on the perspectives of users and societies.
They each might have an opinion about what a helpful \aioracle{} should (and not) do.
Given the potential misalignments between the users, the application developer, and regulators, this tension could appear in the predicate $\phi$.
More importantly, the predicate $\phi$ can be a conjunction of several predicates which means that we can separate helpfulness, through $\phi_1$, and harmlessness, through $\phi_2$.
Yet, whenever one is violated, the predicate $\phi = \phi_1 \wedge \phi_2$ will return false, indicating a failure in the objectives.

\noindent
\textbf{Safety vs. security.} While the model developers ambitiously aim for H\&H,  they are challenged with certain problems: safety and security \cite{taxonomy_of_failures}. 
The goal of \emph{safety} is to prevent the users from unintentional harms. 
Safety guardrails don't protect against a maliciously active attacker. 
Instead, the harm comes from the model's own ``inabilities.'' 

Providing safety is studied by model providers during the reinforcement learning phase, training an RL model that can score \aioracle{} answers on how well they adhere to the two competing objectives. 
Multiple works from (hierarchical) instruction following to constitutional AI \cite{bai2022training, bai2022constitutional, ouyang2022training, mishra2021cross, wallace2024instruction} have emerged in the direction that model developers train the models with safety guardrails in the previous years. 

On the other hand, \emph{security} aims to prevent malicious actors (other than the model itself) to intentionally harm.

\noindent
\textbf{When the loss of safety becomes a security problem.}
\emph{Mismatched generalization} is a vulnerability where a model's safety training fails to be applied to certain inputs.
This happens because the data used to train the safety guardrails of the model is often limited to a few languages with standard natural language text format (e.g. English text and other very common languages). 
The model might have knowledge of less common languages and be able to handle alternative data encodings (such as \textit{Base64}). 
However, its safety guardrails have not been effectively trained to these domains. 
This creates a mismatch between the model's comprehension capabilities and its safety guardrails capabilities. 
This leads an attacker to degrade H\&H by encoding a harmful prompt in a less common format or language, which may not trigger the safety filters trained on plain English, as in \cite{peng2024playing}.

Our objective in this paper is to move the field towards more formal studies. 
In the pipeline given in \autoref{fig:pipeline}, an attacker tries to reach its goals given some capabilities (i.e. which arrow to corrupt in the pipeline). \footnote{If the attacker does not corrupt any arrow, it will be considered as safety problem, rather than the security problem.}

Informally, we will capture the adversaries capabilities in a set called \ATK{}.
\ATK{} will set a flag to indicate if the adversary can see the model, or if it can choose a context, along with a prompt, to play with the \INFER{} phase or choose a context in the \LEARN{} phase. 
We allow these capabilities based on the attacks we have observed during our studies in \autoref{sec:attack_categories}.

\noindent
\textbf{Confidentiality, Integrity, and Availability.} One way to characterize both the safety and security of \aioracle{} is as in traditional information security triad: confidentiality, integrity, and availability.
Even though the defense mechanisms against safety and security problems might differ, CIA triad seems to be still the most meaningful notion to both problems.
In what follows, we discuss how CIA captures the H\&H notions and how they are unified in our security game in \autoref{subsec:formal_games}.

\emph{Confidentiality} focuses on two important principals: protecting the privacy of training data and privacy of data coming from user interacting with the model.
For the former, as discussed in \autoref{subsec:discussing_confidentiality}, we argue that protecting trained data to be revealed to the user is against the usefulness objective.
Such risks should instead be mitigated either by excluding private data from the training corpus through appropriate data‑sanitation procedures, or by fine‑tuning the model on a private dataset whose resulting outputs are accessible only to the authorized users.
The latter covers the leakage of user data  through the result in agentic mode which can be addressed by adding a requirement in the final result. 
Moving ahead, this is captured in our security game through the predicate of $\phi$ in \autoref{alg:security_game}.
    
\emph{Integrity} measurements prevent the responses from being incorrect, unreliable, or tampered in order to useless or to be harm. 
In \aioracle, integrity spans input data integrity, model integrity, and output integrity.
This means protecting against data poisoning (tampering with training data), model attacks (like adversarial examples that manipulate outputs or model weights), and response manipulation (e.g. an attacker altering an AI’s output).
We capture the adversarial capabilities which violate integrity in our security game in \autoref{subsec:formal_games}.

\emph{Availability} makes sure that the system is running and not degraded for the legitimate users.
Availability will also be covered under security game Alg. \ref{alg:security_game} through the predicate $\phi$.


\section{Attack Taxonomy Framework for Language Models}
\label{sec:taxonomy}
As we have seen in previous section, all agentic designs rely on a language model (LM), inherently makes the security of LMs crucial for overall security.
LM attacks have already been subject to various studies, some adapting classic machine learning attack techniques, such as adversarial machine learning techniques to generate adversarial input. 
Some examples of attacks include the Fast Gradient Sign Method \cite{goodfellow2014explaining} and Projected Gradient Descent \cite{madry2017towards}. These are white-box techniques that make small and precise perturbations to the input data to cause the model to produce an incorrect output.
There are other attacks designed to exploit the unique design of the LM architecture. 
We classify the attacks following a multi-dimensional framework, giving a more granular understanding of the vulnerabilities and the possible security improvements. 


\noindent
\textbf{Attack vector.} 
This dimension separates the attacks based on the entry point an attacker uses to accomplish the attack.

\begin{itemize}
    \item \emph{Data-based attacks} succeed when an attacker can perform modifications to some part of the training or fine-tuning data which are used to train the model.
    We capture this by setting \injectLearn{} in \ATK{} in \autoref{alg:security_game}.
    \item \emph{Prompt-based attacks} happen when an attacker can change data sent to model for inference, causing the LM to produce outputs based on the untrusted data. An attacker can modify the user data it can access, or modify the system prompt to ensure system developers guardrails are not considered.
    Our model captures this by setting \injectInfer{} in \ATK{} in \autoref{alg:security_game}.
\end{itemize}

\noindent
\textbf{Attack phase.} This dimension classifies attacks based on the stage of the LM during which the attack is executed. 
\begin{itemize}
    \item \emph{Learning} attacks happen during the model's learning phase. It covers both pre-training and fine-turning rounds in the training process.
    
    \emph{Pre-training attacks} works when the attacker can manipulate the massive dataset used to build the foundation model during the pre-training round. They consist of inserting poisoned data that becomes embedded in the model’s parameters and produces persistent effects on the model’s behavior. Such attacks are difficult to detect as in the initial round the amount of data used is enormous and it is difficult to ensure its quality and security.
    
    \emph{Fine-tuning attacks} target the (smaller) task-specific dataset used to align the model to a specific task or behavior. Such attacks inject malicious data in the fine tuning dataset used to further train the model.
    \item \emph{Inference attacks} happen in the inference phase during live interaction with the model (inference) after it has been trained and deployed.
\end{itemize}

\noindent
\textbf{Adversarial knowledge.}
The capabilities of the attacker depend on the information available about the model, the system and the data used to train the model. 
Following the notation introduced earlier, we distinguish the situations as:
\begin{itemize}
    \item \emph{White-Box Attacks.} In this setting, the adversary has full access to the internals of the model and possibly of the system. 
    The attacker knows the model architecture, weights, training data and has access to the model for inference. 
    An attacker can leverage the model information to craft more advanced attacks based on the model internals, such as based on token gradients. 
    This setting is typical of open-source LM models such as models from the Llama or DeepSeek family.
    Our model captures this by setting \seePipelineInfer{} in \ATK{} in \autoref{alg:security_game}.

    \item \emph{Black-box attacks.} These attacks involve the adversary interacting with the model directly or indirectly via the system, usually via an API for inference. 
    An attacker can only submit instructions to the model and observe the output but has no access to additional information regarding the model. The attack surface is limited and an attacker can only analyze output to malicious prompts to infer weaknesses. Models susceptible to these attacks in this category include the ones provided by Anthropic (i.e. Claude).
    In this case, \seePipelineInfer{} is not set in \ATK{}, but \textsf{black\_box} is in \ATK{} in \autoref{alg:security_game}.
\end{itemize}

Our games do not model the grey-box type of adversarial knowledge where the attacker has a restricted internal visibility of the model.

\noindent
\textbf{Adversary's source.} The origin and role of the attacker is critical for correctly classifying threat models.

\begin{itemize}
    \item \emph{User}: an adversary is the end user interacting with the system for e.g. creating harmful or illegal outputs.
    \item \emph{Third-party}: the adversary is a third party user that poisons data sources used by the application.
    \item \emph{Supply-chain}: the adversary has access to the model training data and tries to poison it.
\end{itemize}

\noindent
\textbf{Attack persistence.}
The universality of an attack makes it more dangerous as it is easier and more common to reproduce and exploit. This dimension measures this metric by quantifying the longevity of the attack’s effect.
\begin{itemize}
    \item \emph{Transient attacks} are limited to a single interaction or session and therefore does not halter the model’s state or permanent behaviors.
    \item \emph{Persistent attacks} are long-lasting and affect the model’s future behaviors. Those attacks try to halt the model’s parameters and internal state, causing possible differences in the model’s future outputs.
\end{itemize}

\section{Applying the Taxonomy: Attack Categories}
\label{sec:attack_categories}

In this section, we analyze three major categories of LM attacks. 
For each category we specify the attack vector, attack phase, adversarial capabilities (via the \ATK{} set) and source, attacker goals (which games the adversary tries to win) and attack persistence.

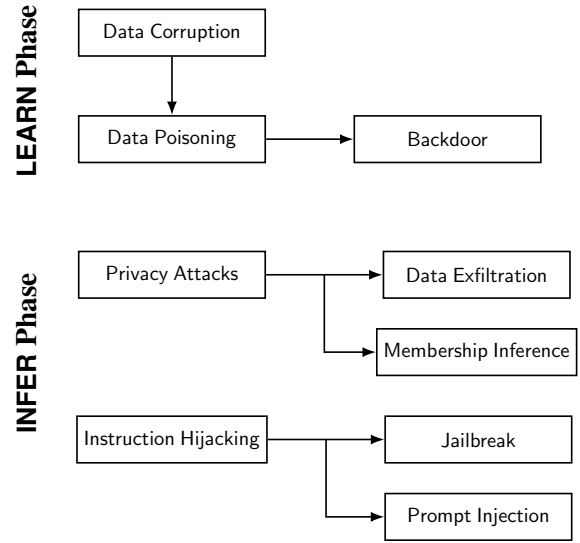
\begin{figure}[t]
  \centering
  \resizebox{0.9\columnwidth}{!}{%
  \begin{tikzpicture}[
    node distance=1cm,
    >=Latex,
    attack/.style={
      rectangle,
      draw,
      thick,
      minimum height=8mm,
      minimum width=32mm,
      align=center
    },
    phase_label/.style={
      font=\Large\bfseries,
      rotate=90,
      anchor=center
    }
  ]

    \node[phase_label] at (0, -1) (training_phase) {\LEARN{} Phase};
    \node[phase_label] at (0, -5.5) (inference_phase) {\INFER{} Phase};

    \node[attack] (corruption) at (2.5, 0) {$\mathsf{Data~Corruption}$};
    \node[attack, below=of corruption] (poisoning) {$\mathsf{Data~Poisoning}$};
    \node[attack, right=1.5cm of poisoning] (backdoor) {$\mathsf{Backdoor}$};

    \draw[->, thick] (corruption) -- (poisoning);
    \draw[->, thick] (poisoning) -- (backdoor);

    \node[attack, below=1.5cm of poisoning] (privacy) {$\mathsf{Privacy~Attacks}$};
    \node[attack, below=2cm of privacy] (hijacking) {$\mathsf{Instruction~Hijacking}$};

    \node[attack, right=2cm of privacy] (exfiltration) {$\mathsf{Data~Exfiltration}$};
    \node[attack, below=0.5cm of exfiltration] (membership) {$\mathsf{Membership~Inference}$};
    \node[attack, right=2cm of hijacking] (jailbreak) {$\mathsf{Jailbreak}$};
    \node[attack, below=0.5cm of jailbreak] (prompt) {$\mathsf{Prompt~Injection}$};

    \coordinate (privacy_fork) at ([xshift=1cm]privacy.east);
    \draw[->, thick] (privacy) -- (privacy_fork) |- (exfiltration.west);
    \draw[->, thick] (privacy_fork) |- (membership.west);

    \coordinate (hijacking_fork) at ([xshift=1cm]hijacking.east);
    \draw[->, thick] (hijacking) -- (hijacking_fork) |- (jailbreak.west);
    \draw[->, thick] (hijacking_fork) |- (prompt.west);

  \end{tikzpicture}
  }
    \caption{An overview of LM attack categories grouped by the model's phases. Arrows denote progressive specialization in the type of attacks.}
  \label{fig:llm_attacks_compact}
\end{figure}

A comparative overview of the discussed attacks in this section is in \autoref{table:attack_categories}, categorizing them according to our multi-dimensional framework.
We note that the set of attacks we include here is not the complete list.
A new attack can emerge, yet our framework should be able to capture them.

\begin{table*}[t]
\centering
\begin{tabular}{|c|c|c|c|c|c|c|}
\hline
 & \textbf{Attack} & \textbf{Attack}  & \textbf{Adversarial}  & \textbf{Security}  &  & \textbf{Adversary}  \\
 & \textbf{Vector} & \textbf{Phase} & \textbf{Knowledge} & \textbf{Goals} & \textbf{Persistence} & \textbf{Source} \\
\hline
\textit{Prompt Injection} & Prompt-based & Inference & \shortstack{White-Box \\ Black-Box \\ Gray-Box} & Confidentiality & Transient & \shortstack{User\\Third-Party}\\
\hline
\textit{Jailbreak} & Prompt-based & Inference & \shortstack{White-Box \\ Black-Box \\ Gray-Box} & Integrity & Transient & User \\
\hline
\textit{\shortstack{Data \\ Exfiltration}} & Prompt-based & Inference & Black-Box & Confidentiality & Transient & User \\
\hline
\textit{\shortstack{Membership \\ Inference}} & Prompt-based & Inference & Black-Box & Confidentiality & Transient & User \\
\hline
\textit{\shortstack{Data \\Poisoning}} & Data-based & Learning & - & \shortstack{Integrity \\ Availability} & Persistent & Supply-Chain \\
\hline
\textit{Backdoor} & Data-based & Learning & - &  Integrity & Persistent & Supply-Chain \\
\hline
\end{tabular}
\caption{Cross-category attack comparisons.}
    \label{table:attack_categories}
\end{table*}


\subsection{Data Corruption Attack}
This attack category covers the attacks which are corrupting or manipulating the corpus during LM's training (i.e. $\injectLearn \in \ATK$) to influence the model behavior. 

Data corruption is a \underline{data-based} attack during the \LEARN{} phase (\underline{pre-training} or \underline{fine-tuning}) with \ATK{} set containing $\injectInfer$. The attack can be formalized as a security game, with the attacker intercepting the data from \truth{} and modifying it to create a poisoned context / training set for the \LEARN{} algorithm.

The attack success is measured in after the training is completed, with the attacker measuring if the model inference output is not aligned with H\&H goals. The attacker does \emph{not need knowledge} of the model internals, just of the corpus the model might use.
The goal of an attacker is to violate the \underline{integrity} of the model, by creating fault model behaviors by poisoning corpus later used in the training process.
Those attacks are permanent, as the model is modified by including a persistent anomalous behavior.

This attack has two sub-categories: data poisoning and backdoor (trojan) attacks as detailed in \autoref{app:data_corruption}.

\subsection{Instruction Hijacking Attacks}
Instruction hijacking aims at manipulating the model's behavior, making it deviate from its intended instructions. 
The attacker could be a malicious user prompting directly to the LM or a 3rd-party attacker injecting information into the LM.

This category of attacks takes place during the \underline{\INFER{}} phase and consists in having an attacker craft a malicious \prompt{} that makes the model produce an output that deviates from the LM’s intended safety guidelines, violating the system integrity. 
Attacks from this category are \underline{prompt-based attacks} that are \underline{non-persistent}: model unsafe behavior is only limited to the single interaction or session. Depending on the way the attacker crafts the malicious \prompt{}, the attacker might require more or less knowledge of the model. 

Formally, instruction hijacking is an instance of the security game \autoref{alg:security_game} and the adversary \ATK{} set contains $\injectInfer$. 
The attack is successful if the model generates a \result{} that violates the safety guidelines in the \truth{} set.

These attacks happen as LMs handle heterogeneous data that contains both data and instructions to follow, however they are not able to distinguish between the instructions coming from the \context{} and those coming from \prompt. 
We detail this category with prompt injection and jailbreak attacks in \autoref{app:instruction_hijacking}.

\subsection{Privacy Attacks}
Models are usually trained on vast amounts of corpus (either public or private). 
Such data might contain personal or proprietary information that should not be retrievable via model inference, causing privacy leakage problems. 
The attacker is a malicious user interacting with the model to violate the \underline{confidentiality} property by extracting information that should have not been revealed.

The attacks in this category happen during the \underline{\INFER{}} phase, and the attacker does \underline{not require additional information} regarding the model internals (except potential information on the corpus the model is trained on). 
As the attacks in this category are achieved during the model inference, those are \underline{prompt based attacks}. 
Privacy attacks do not alter the model itself therefore the attacks in the category are \underline{temporary} and limited to the malicious session. 
These attacks happen as LMs, due to the vast amount of corpus they process, lack proper data sanitation during training and other rigorous privacy protection mechanisms, allowing private information retrieval.
Data exfiltration and membership inference attacks are the most common sub-category of privacy attacks as detailed in \autoref{app:privacy}.
\section{Agentic Systems and Access Control} \label{sec:aacm}
The World Economic Forum defines the agentic systems as ``autonomous systems that sense and act upon their environment to achieve goals” \cite{WEF_agents, taxonomy_of_failures}. 
In our work, we will consider agents as \aioracle{} which could be seen as a machinery where the output is a program to be executed.
This execution may interact with external tools which provide additional input and output channels.

The security community has increasingly sought to define \textit{agentic security}. 
It is not an entirely separate concept from traditional computing security, but it requires a paradigm shift.
Agentic AI systems integrate LMs for reasoning, planning, and task execution. 
Beyond inherent LM vulnerabilities, agentic systems expand the attack surface.
This is due to three factors: reliance on LM outputs for planning; integration with external tools via function calls; and access to environments containing user data and external sources. 
Several methods for safeguarding against this emerging attack surface have been introduced in the literature. 

\noindent
\textbf{Agentic Access Control Mechanisms.} While benchmarks \cite{liu2024formalizing, debenedetti2024agentdojo, zhang2024agent, evtimov2025wasp, chao2024jailbreakbench} provide an empirical metric to analyze and quantify the security of defense mechanisms, in this section we leverage our multi-dimensional attack taxonomy (\autoref{sec:taxonomy}) and attack categorization (\autoref{sec:attack_categories}) to evaluate agentic access control mechanisms (\aacm{}).

The goal of an \aacm{} is to determine what the agent, who is executing a program on behalf of the user, can/not do.
These mechanisms are brought to bring security and safety to the AI systems in a way that the user who is using the agent does not get harmed.
Typically, such controls are brought through some defined policies that enforces an agent to follow.
On the other hand, \aacm{} should not make the AI-based systems not useful while focusing on the security and safety.

There are two folds of the \aacm{} designs: (1) train the models to follow some policies during the \LEARN{} phase and (2) put additional guardrails for the \result{} of models before they execute any task.
In the remainder of this section, we elaborate on both components and explain how each of these ideas integrate in our formalism results in \autoref{subsec:integrate_agentic_approaches}.

\subsection{Model Training Principles}
Before an agent can use external tools and adhere to certain rules, the underlying LM is trained to be more controllable and aligned with a core set of principles. 
In the latest years, two ways on how LMs adhere to given instructions and safeguards have emerged: \textit{Instruction Following} and \textit{Constitutional AI}.

\subsubsection{Instruction Following} \label{subsec:instruction_following}
The instructions the model has to follow are heterogeneous: they include data from multiple sources, past conversations and multiple conflicting instructions. 
Early work \cite{ouyang2022training, mishra2021cross} focused on showing that, by fine tuning models with instructions-response pair, the model can be trained to follow developers policies and accomplish user goals more effectively. 
Another work \cite{wallace2024instruction} studied an evolution of the instruction following principle known as Hierarchical Instruction Following to improve the overall reliability and security.
It is based on the idea that significant vulnerabilities, leading to instruction hijacking attacks, arise if the LM does not distinguish between prompt data sources: user prompt, system instructions, and tool output (as a \context). 
Hierarchical instruction training addresses the problem by fine tuning the LM to prioritize instructions based on their sources, establishing in the model internal weights and thinking process a source hierarchy: system instructions then user instructions then tool output.


\subsubsection{Constitutional AI (CAI)} \label{subsec:cai}
CAI, introduced in \cite{bai2022constitutional}, is another way to train an LM to follow a set of principles: the \textit{constitution}.  
The specific term \textit{constitution} meant to highlight the significance of harmlessness of any model. 
The main idea is to start from a model which is trained to be the most helpful to the user, i.e. answers all the questions regardless of the harm it may cause to the users or the others, and retrain it through the constitution, a set of generic behaviors and rules to follow, to learn how to be harmless.
CAI includes several \aioracle{s}: main LM which updates itself with a ``critique'' LM and the reinforcement learning model without human feedback.

First, the model learns to follow the constitution principles through two-step process: \textit{critique} and \textit{revise}. 
The model at first generates a response for an initial prompt.
Critique step requests the LM to be critical about its own response with certain checks such as discrimination, age appropriateness, and/or legal implications of the tasks prompted by the users. 
The critique criteria stems from a principle extracted from a constitution which is written by the model developer themselves and provided to the model to create a response from this principle. 
The revision step requires the LM to revisit the original response according to the critique step feedback.
These steps on a large number of prompts are repeated to create a supervised learning dataset. 
Finally the LM is further fine-tuned using \reinforce{} algorithms and the previous reward model as the reward function. 
This process makes the answer produced by the LM to be the one best fits the constitution.

While this approach shows substantial improvements for the model harmlessness, it comes with the cost of decreasing the helpfulness and, for smaller models, the additional synthetic data can causes model collapse as shown with \texttt{Llama 3-8B} in \cite{zhang2025constitution}.
An approach similar to CAI is applied in multiple model in use today. 
For example Apertus \cite{hernandez2025apertus}, a model developed by the Swiss AI Initiative, was aligned using a set of principles derived from the Swiss constitutional values. 

\subsection{Model Inference Principals}
\label{sec:architectural_frameworks_agentic_ai}
Model level defenses, focusing on making the model itself more robust (such as our dual construction as described in \autoref{subsec:formal_games}), work well in theory. However, this approach improves general safety but cannot be considered a universal defense, as the model remains vulnerable to more sophisticated attacks targeting its architecture and internals \cite{pandya2025may}.
To address this limitation, another defense paradigm studied: system level defenses. 
This approach treats the model as an untrustworthy black box and builds a secure environment around it.

\begin{table*}[t]
\centering
\begin{tabular}{|c|c|c|c|}
\hline
\textbf{Framework} & \textbf{Execution Model} & \textbf{Core Concept} & \textbf{Standard Security Concept}\\
\hline
FIDES \cite{costa2025securing} & Deterministic & \shortstack{Data tracking and label\\ propagation via taint-tracking} & \shortstack{Information flow control \\ Lattice-based Access control}\\
\hline
Progent \cite{shi2025progent} & \shortstack{Deterministic enforcement\\ of non-deterministic policies} & Privileged tool access control & \shortstack{Principle of least privilege\\ Access control} \\
\hline
Conseca \cite{tsai2025contextual} & \shortstack{Deterministic enforcement of \\ non-deterministic policies} & \shortstack{Just-in-time and context-aware\\ policy generation} & Access control \\
\hline
\shortstack{CONTROL-\\VALVE} \cite{jha2025breaking} & \shortstack{Deterministic CFG check \\and non-deterministic LM judge} & \shortstack{Securing agent interactions\\ and executions} & control-flow integrity \\
\hline
\end{tabular}
\caption{Comparison of model inference principals for \aacm{}.}
\label{table:framework_comparison_concise}
\end{table*}

\subsubsection{FIDES} \label{subsec:fides}
\textbf{Core concept.} As in traditional software security, information-flow-control can be used to provide security guarantees to AI agents and to track each piece of data in use by the system and avoid or limit the use of untrusted data.
In this spirit, FIDES introduces an access control mechanism by attaching confidentiality and integrity labels to all data processed by the agentic system \cite{costa2025securing}.

\noindent
\textbf{Framework enforcement method.} FIDES allows deterministic policies that analyze the confidentiality and integrity of the data to be enforced.
The system proposes a data tracking mechanism, a lattice based access control with clear ordering of privileges introduced in \cite{denning1976lattice}, to help the system propagation of integrity and confidentiality labels.  
FIDES also defines variables to contain the output of the external tools whose content cannot be used for data dependent actions without further secure processing.

\noindent
\textbf{Limitations and tradeoffs}
FIDES uses deterministic policies to allow certain information flows and properties for a tool call to happen. 
The policies are manually assigned to the tools and their outputs.
An \aacm{} integrated with the FIDES framework would be a series of policies and tools, connected to an LM. 
The tools are denoted by confidentiality and integrity tags that the policies deterministically enforce during the framework runtime i.e. whenever a task the user provided is being solved. 
If the task solving plan fails to respect the policies, then the system \emph{tries} to propose a different plan or halts.
It is not clear how often the system ends up halting.

\subsubsection{Progent} \label{subsec:progent}
\textbf{Core concept.} \textit{Progent} \cite{shi2025progent} is a privilege control framework for securing \aacm{}, implemented through a domain-specific language based on JSON schema. 
The core focus addresses the problem of \textit{over-privileged tool access}, which enables various attacks to succeed, by restricting agents to performing only tool calls necessary for user tasks.

\noindent
\textbf{Framework enforcement method.} Progent provides fine-grained access control through conditional allow/forbid rules, enabling users to define which tools are permitted or forbidden, specify fallback actions when tool calls are blocked, and implement dynamic policy updates that adapt to changes in an agent state. 
The authors also propose \textit{Progent-LM}, an LM-assisted approach for automated policy generation on a per-user-query basis, aligning with the principle of least privilege.

\noindent
\textbf{Limitations and trade-offs.} A primary limitation is its mere focus on the tool call phase, making attacks targeting text outputs remain unaddressed. 
Another limitation is that the deterministic security guarantees depend heavily on the correctness and completeness of manually-written policies. 
While the authors provide validation tools (a type checker and a condition overlap analyzer) and explore LM-assisted policy generation as a potential mitigation, manual policy specification remains error-prone.

Moreover, the LM-assisted dynamic policy update mechanism introduces potential vulnerabilities through reliance on prompt engineering. 
Although the authors employ a two-step mitigation process to limit this risk by requiring attackers to compromise two separate prompts (prompt for policy update check and prompt for performing policy update), the defense remains fundamentally dependent on the LM correctly filtering malicious instructions from untrusted tool outputs. 
When tool results contain adversarial prompts, the update mechanism may inadvertently relax security constraints. 
We argue that implementing a \textit{Dual-LM pattern}~\cite{willison2023dual} where a separate model evaluates policy updates in isolation from the primary agent LM would significantly improve robustness by providing architectural separation rather than relying solely on prompt engineering.

\subsubsection{Conseca} \label{subsec:conseca}
\textbf{Core concept.} \textit{Conseca} \cite{tsai2025contextual} explores contextual security for LM agents by generating dynamic, context-aware policies. 
The authors argue that manually pre-written static policies, although fine-grained, cannot account for every possible context, leading to either over-restriction or over-permissioning depending on the situation.

\noindent
\textbf{Framework enforcement method.} Conseca abstracts agents into two components: a ``planner'' that processes user requests and proposes actions to execute, and an ``executor'' that runs these actions and interfaces with external tools.
The planner is a trusted component.
Upon receiving a user task, Conseca generates just-in-time security policies based on trusted context, then deterministically enforces these policies by evaluating each proposed action by planner before execution.
This is how we can formalize the agents in terms of \aioracle{s}.
The \aioracle{} is the planner and the environment the executor.

\noindent
\textbf{Limitations and trade-offs.} A significant limitation of Conseca is its vulnerability to \textit{adversarial planning attacks}, where a compromised planner strategically evades security policies. 
Since the planner processes all (un/trusted) context while Conseca is assumed to see trusted context, a planner compromised via prompt injection can model Conseca's policy generation and propose sequences of individually benign-looking actions that collectively achieve malicious goals. 
Unlike traditional access control systems where the policy enforcer is a separate entity with distinct information access, both the planner and Conseca in this architecture are LM-based components operating on overlapping information, enabling the planner to reason about and work within Conseca's constraints. 
While the authors acknowledge that Conseca currently checks only individual actions rather than trajectories, this limitation is particularly problematic when the planner itself can anticipate which action sequences will pass policy checks. 
Finally, Conseca's approach of using an LM-based policy generator to defend against prompt injection attacks on the planner LM creates a fundamental ``LM-to-secure-LM'' paradox, where the defense mechanism remains vulnerable to the same class of attacks, essentially adding another attack surface rather than eliminating it. 
This design implicitly assumes the policy generator is a safe \aioracle{} component, which undermines the framework's security guarantees.

\subsubsection{Controlvalve} \label{subsec:contravalve}
\textbf{Core concept.} \textit{Controlvalve} \cite{jha2025breaking} is a task-agnostic defense designed to protect agentic systems by applying Control-Flow Integrity (CFI), similarly to the standard CFI approach to secure programming languages. 
The framework secures the agent's interactions and executions via a two stage process. 

\noindent
\textbf{Framework enforcement method.} In the first planning stage an LM is used to generate two security specifications: a control flow graph (CFG) and a set of natural languages rules. 
The CFG consists of the legitimate paths of agent function invocations.
The rules are edge-specific and contextual and specify the conditions for each interaction to be permitted. 
In the second stage the agent's actions are executed and the defined guardrails are enforced. 
Each action must first pass a CFG deterministic check and then an LM judge to ensure the action content and context obey to the edge-specific rules.

\noindent
\textbf{Limitations and trade-offs.} The security of the system highly depends on the quality of the LM generated security CFG and rules. 
While during the planning phase the information used to generate the artifacts are trusted, the LM may fail to generate a sufficiently strict rules (due to over-approximation), leaving a security hole in the system for the overall task's lifecycle.
\section{Our Formalism for \aioracle{s}} \label{sec:formalism}

An \aioracle{} works in two phases: \LEARN{} and \INFER{} where they might be overloaded as algorithms.
That is $\model \gets {\LEARN}(\textsf{data})$ where \data{} is extracted from a set denoted as \truth{} such that $\data \gets \genEntryForPhase(\truth)$.
\truth{} is the inaccessible set defining what exists about ``life, the universe, and everything'' (Part III of the Hitchhiker's Guide to The Galaxy).
\data{} is generated in nature from this set and \model{} learns from this data.
And, \INFER{} inputs $(\prompt, \context, \model)$ and outputs \result{}.

We also need the notion of a predicate $\phi$ which defines what is correct, useful, and harmless. 
The predicate $\phi$ is \textbf{not} implementable.
The \truth{} is common to every application, while $\phi$ depends on the application.

\subsection{Notations}
\textbf{System components.} Before performing our analysis of \aioracle{} attacks, we first give a clear description of the (agentic) system components: \textit{\aioracle{}, data, tools, and participants}.

\begin{itemize}
    \item An \aioracle{} has two phases defined with a set of algorithms as given in \autoref{fig:pipeline}. 
    \item \genEntryForPhase{} takes some information from existing sources, called \truth{}, in an informal manner, then formats it following this algorithm which must be implementable.
    The output is called \data{}.
    Intuitively, the set \truth{} is the set of all available data from which we extract the \data{} to be learned.
    Then, \LEARN{} algorithm uses \data{} to output the model.

    \item Tools are external applications developed by 3rd parties.
    \item We identify two main participants:
    \emph{User} is the individual interacting with the agentic application to solve a task.
    \emph{Attacker} $\adversary$ could be the user or the third party (e.g. external tool or data provider) interfering the system.
\end{itemize}
 
\noindent
\textbf{Interaction data.} The interaction starts from the user and the system that provides data that is shared with the model provider, which in turn responds with an output.
We have the following data from \INFER{} phase.

\begin{itemize}
    \item \prompt{} refers to an untrusted input provided by the user, it can be a direct string the user prompted or a fixed string for an action the user might wish to do. 
  
    \item \context{} denotes the data which the application or system provides the model to support the execution of the user task from the \prompt{}. 
    It may include the system prompt (i.e. instructions provided by the (agentic) system developers that define the objectives and constraints for the interaction with the user); the external data stored and retrieved from tool execution; or the tool output which returned from an invoked tool call.
      
    \item \result{} is the sequence of tokens generated by the model given the input prompt. The output might be a simple string returned from the application to the user or might also contain keywords to call external tools.
\end{itemize}

The predicate $\phi$ defines the set of accepted (\prompt, \context, \result) triplets which are considered as ``valid'' (i.e. aligned with its H\&H goals) from $\INFER(\prompt, \context, \model)  = \result$ interactions. A true (resp. false) predicate returns 1 (resp. 0).

\subsection{Definitions}
\textbf{\aioracle.} The learning process \LEARN{} is highly dependent on a function called \genEntryForPhase{} which extracts data from the set \truth{}.
The input in phase \LEARN{} is the result of $\genEntryForPhase(\truth)$ which will sample from existing data sources some elements to inject.
We define the \aioracle{} as a set of algorithms where the learning phase is a pair of algorithms, $(\genEntryForPhase, \LEARN)$, run through $r$ rounds. 
We split the learning phase into rounds to model the attacks on specific rounds. 
Contrarily, \INFER{} could also be iterative but there is no attack on specific iterations.
This must be due to that rounds during learning are more complex than \INFER{} iterations.
More formally:

\begin{definition}{}
For an $r$-round of learning phase, an \aioracle{} is defined by a set of $2r+1$ algorithms $((\genEntryForPhasei, \LEARN_i)_{i \in \{1, \ldots, r\}} , \INFER{})$.
For each $i$, $\genEntryForPhasei(\truth)$ returns a subset of \truth{} and this subset is used in $\LEARN_i$.
\end{definition}

Given that an \aioracle{} is a set of algorithms with two phases, in our security definition, we will allow the adversary to interact with \aioracle{} one of these phases depending on the given capabilities.

\noindent
\textbf{Completeness.} 
We define the completeness of an \aioracle{} relative to a set \truth{} and for a predicate $\phi$. 
The completeness is a notion which is characterized in the absence of an adversarial behaviour. 
Thus, the definition is meant to have no malicious activities, that is the ``adversary" does not have the power to interfere with the \aioracle{}. 
Instead, the game itself calls the algorithms of \aioracle{} when needed.

The predicate $\phi$ says if a result from the inference phase is aligned with the design goals: it is correct, useful, and harmless \footnote{Note that the predicate covers both H\&H objectives, but can be separated easily into sub-predicates.}.
Furthermore, since the set \truth{} is typically gigantic, algorithms do not exactly take it as input but we assume that \truth{} is a structured data source and that algorithms have random read access to any element of it. 
In the following completeness definition, we just want to generate a (\prompt{}, \context{}) pair by any algorithm $\mathcal{B}$.

\begin{definition}
Given a set \truth{}, an \aioracle{} for a predicate $\phi$ is $p$-\emph{complete} if for any probabilistic polynomial-time algorithm $\mathcal{B}$, the following game returns 1 with prob. at least $p$.
\begin{algorithm}[h!]
  \begin{algorithmic}[1]  
    \State $\model \gets \emptyv$
    \For{$i \gets 1$ \textbf{to} $r$}
      \State $\data_i \gets \textproc{\genEntryForPhase}_i(\truth)$
      \State $\model \gets \LEARN_i(\model,\, \data_i)$
    \EndFor
    \State $(\prompt,\context) \gets \mathcal{B}(\truth)$
    \State $\result \gets \INFER(\prompt,\context,\model)$
    \State \Return $\phi(\prompt,\context,\result)$
  \end{algorithmic}
\end{algorithm}
\end{definition}

Here, $\mathcal{B}$ is not given the model itself but can generate more data from the $\truth$ set. 
This will help us to  capture the composition of different \aioracle{s} in the security games.

\noindent
\textbf{Decisional vs computational problem.} We define two types of problems which \aioracle{} tackles to solve: a decisional and a a computational problem. 
We note that any problem is a computational problem and that decisional problems are particular cases where the \result{} can be either an ``accept'' or ``reject''.
The source of inspiration of this separation is modern cryptography where we have computational and decisional Diffie-Hellman problems with different difficulties \footnote{Decisional Diffie-Hellman is an easier problem due to the fact that the adversary is given a hint compared to the computational version.}.
Having an \aioracle{} for decisional problem corresponds to designing an \aioracle{} which tells whether $(\prompt, \result)$ is a correct pair for a given \context{} (i.e. the \result{} is a hint to output an acceptance bit) whereas the computational problem is that given a \prompt{} and a \context{}, the oracle should compute a valid \result{} (with no hint).
More formally, a predicate $\phi$ for a computational problem defines another predicate $\phi'$ for the companion decisional problem by $\phi(\prompt, \context, \result) \Leftrightarrow \phi'((\prompt, \result), \context,$ ``\accept'').

\subsection{Example of Agentic System as an \aioracle} \label{subsec:aioracle_example}
We now describe a ``simple'' task which clarifies the internal workings of \aioracle{} as an agent.
The user prompts to the \aioracle{} $\prompt_1 = \text{``summarize the unread emails from my inbox.''}$
The output of the call $\result_1 = \aioracle.\INFER(\prompt_1, \bot, \model)$ is a code similar to \autoref{alg:simple_agent}.

\begin{algorithm}[h!]
  \caption{\textsf{Agent}($\result_1$)}
  \label{alg:simple_agent}
  \begin{algorithmic}[1]  
    \State \text{Call the email client as an external tool}
    \State \textsf{tool\_output$\mathsf{_1}$} $\gets \text{tool(unread emails)}$
    \State $\prompt_2 \gets$ ``summarize what follows: \textsf{tool\_output$\mathsf{_1}$}''
    \State $\context_2 \gets \prompt_1$
    \State \text{Call} $\aioracle{}(.,.,\model)$ \text{ as an external tool}
    \State $\result_2 \gets  \textsf{tool}(\prompt_2, \context_2)$
    \State \text{send } $\result_2$ \text{ to the user}
  \end{algorithmic}
\end{algorithm}

The proposed code itself in $\result_1$ is generated by the \aioracle{} from the user's initial prompt and it satisfies $\phi$ because it corresponds to what the user requested. 
Once the code in $\result_1$ is executed by the system, it will retrieve the unread emails and make a call to itself (line 5) to summarize the emails.
The final email summary, $\result_2$, will be returned by this execution.
The second execution of \aioracle{} (in line 5) should also produce $\result_2$ such that $\phi(\prompt_2, \context_2, \result_2)$ is true.
For instance, it should resist to prompt injection attacks (\autoref{subsubsec:prompt_injection}) by not interpreting email content as action orders.

\subsection{Discussions On Confidentiality} \label{subsec:discussing_confidentiality}
\textbf{Confidentiality vs utility.} Confidentiality is usually defined as what a malicious user could learn from interacting with the model.
There exist many different ways to define it. 
We give the DPD model from \cite{salem2023sokletprivacygames, salem2023sp} with our notations as a typical example. 
In this framework, there is a single round of learning ($r=1$) and a set $\data$ which is an unordered set of $n$ elements. 
The adversary builds the dataset with two options $\data'\cup\{z_0\}$ or $\data'\cup\{z_1\}$ which differ by only one element. 
The model learns from one of these options and the adversary must guess which one from the model.
This perfectly models the membership inference attack from \autoref{subsubsec:inference_atk}.

\begin{figure}[t]
  \centering
  \begin{algorithm}[H]
    \caption{\textsf{DPD}$(n)$}
    \label{alg:confidentiality-game}
    {\small
    \begin{algorithmic}[1]

      \State flip a coin $\bcoin$
      
      \State $(\data',z_0,z_1) \gets \adversary'(n, \truth)$
      \State \textbf{if } $|\data'|\neq n-1$ \textbf{ then } \Return 0
      \State $\data \gets \data' \cup \{z_b\}$

      \State $\model \gets \LEARN(\emptyv,\ \data)$

      \State $\bprime \gets \adversary{(\data', z_0, z_1, \model, \truth)}$
      \State \Return $(\bcoin = \bprime)$

    \end{algorithmic}
    }
  \end{algorithm}
  \caption{Algorithmic description of differentially private distinguishability (\textsf{DPD}) game.}
  \label{alg:dpd}
\end{figure}

However, this framework does not discuss about completeness.
And, there is a clear tension between confidentiality, where we do not want to learn from a specific data element, and completeness, where we do want to learn every data element.
To illustrate this, we consider a simpler classifier use case with a set of images of cats and dogs.

The tension is that if we have a \textsf{DPD}-secure \aioracle{} (i.e. the classifier), then \aioracle{} cannot distinguish a cat from a dog which undermines the utility of an \aioracle{}.
We show this claim formally in \autoref{appendix:sec_utility}.

\noindent
\textbf{Confidentiality vs integrity.} The tension between confidentiality and utility comes from the fact that \textsf{DPD} security treats all data elements from corpus as confidential.
In practice, it is an overkill because we do not necessarily need and want to protect all data elements. 
Pragmatically speaking, we have three options: the learning phase treats confidential data differently, or eliminate the confidential data from corpus before learning, or add another restriction in $\phi$ in a way that the \result{} does not include any confidential information. 
In all cases, we need to identify confidential information.
Again, identifying confidential information is yet another AI classification problem or a decisional problem where the objective is characterized by a predicate $\phi'$.

The confidentiality of user data is more severe in agentic systems.
However, we need a proper $\phi$ to characterize if the \result{} would leak confidential user data or not.
Consequently, all confidentiality problems reduce to ensuring that the triple $(\prompt, \context, \result)$ satisfies $\phi$.

\subsection{Security Game} \label{subsec:formal_games}
In this section, we move towards more precise formalism for security issues by games played by a malicious adversary.
We model the game with two sets: \ATK{} and \truth{}.
The set \ATK{} indicates what the attackers capabilities are and captured with four labels: $\seePipelineInfer, \injectLearn, \injectInfer, \blackbox$. 
If $\seePipelineInfer$ is in \ATK, it means that the adversary can see the output from \LEARN{} phase.
If $\injectLearn$ is in \ATK, then the adversary can add a chosen input $\context$ in the rounds of \LEARN{} phase.
If $\injectInfer$ is in \ATK, then the adversary can corrupt the \data{} in \INFER{} phase.
If \blackbox{} is  $\in \ATK$, then the adversary can play with \INFER(.,., \model).
The white-box model corresponds to $\seePipelineInfer_r \in \ATK{}$.
Finally, the adversary is choosing the prompt made by the user.

\noindent
\textbf{Security Game.}
Our security game captures confidentiality, integrity and availability. 
One simple version of our security game is defined in \autoref{alg:simple_security_game} in \autoref{fig:simple_security} with \ATK{} defined with two flags which capture the white-box attacks (i.e. the adversary can see the final output \model{} via \seePipelineInfer{} flag) with prompt and context injection capabilities (i.e. \injectInfer{} flag is on and adversary chooses both \prompt{} and \context{} to the \INFER{} phase).
The full security game which captures all attacks is defined in \autoref{alg:security_game} in \autoref{fig:security}.
The goal of the adversary, i.e. the winning condition, is to make the predicate $\phi(\prompt, \context, \result)$ not satisfied.
The game does not allow trivial wins.
For example, if the $\injectLearn_i \in \ATK$ for every $i$, i.e. the attacker can change $\data{}_i$ to train the model to learn that elephants can fly and ask in the prompt if elephants can fly.
We check if an attack is trivial by verifying a predicate $\psi(\textsf{trace})$ which is computed on the trace of the game execution.
Notice that the completeness game corresponds to the security game with $\ATK = \{\injectInfer \}$ (without output flipped) where $\mathcal{B}$ playing as an adversary.

If we focus on attacks where $\injectLearn_i \not\in \ATK$ for every $i$, we make the adversary passive during the learning phase.
Such passive adversary will not have trivial attacks on the system.
$\Psi$ is always false.
In other cases, $\Psi$ should be based on the pairs $\data_i \to \data'_i$ (as it is changed by $\adversary$ in each round) and a triplet $(\prompt, \context, \result)$.

The advantage of the adversary is defined with $\advantage = \Pr[\SecurityGame \to 1] - (1-p)$.
The baseline is a completeness adversary making the model return an invalid result with probability $(1-p)$.

\begin{figure}[t]
  \centering
  \begin{algorithm}[H]
    \caption{\SecurityGame$(\ATK, \truth)$}
    \label{alg:simple_security_game}
    {\small
    \begin{algorithmic}[1]

      \State assert $\ATK = \{\seePipelineInfer_r, \injectInfer\}$
      \State $\model \gets \emptyv$

      \For{$i = 1$ \textbf{ to } $r$}
        \State $\data_i \gets \genEntryForPhasei(\truth)$
        \State $\model \gets \LEARN_i(\model,\ \data_i)$
      \EndFor

      \State $(\context,\ \prompt) \gets \adversary(\model, \truth)$

      \State $\result \gets \INFER(\prompt,\ \context,\ \model)$

      \State \textbf{if } $\phi(\prompt,\ \context,\ \result)$ \textbf{ then } \Return $0$
      \State \Return $1$

    \end{algorithmic}
    }
  \end{algorithm}
  \caption{Algorithmic description of security game with flags $\seePipelineInfer_r$ and \injectInfer{}. }
  \label{fig:simple_security}
\end{figure}

\begin{figure}[t]
  \centering
  \begin{algorithm}[H]
    \caption{\SecurityGame$(\ATK, \truth)$}
    \label{alg:security_game}
    {\small
    \begin{algorithmic}[1]

      \State $\model \gets \emptyv$
      \State $\state \gets \emptyv$

      \For{$i = 1$ \textbf{ to } $r$}
        \State $\data_i \gets \genEntryForPhasei(\truth)$
        \State \textbf{if } $\seePipelineInfer_{i-1} \in \ATK$ \textbf{ then } $\view \gets \model$ \textbf{ else } $\view \gets \emptyv$
        \State $(\data'_i, \state) \gets \adversary(\state, \view, \truth)$
        \State \textbf{if } $\injectLearn_i \not\in \ATK$ \textbf{ then } $\data'_i \gets \data_i$
        \State $\model \gets \LEARN_i(\model,\ \data'_i)$
      \EndFor

      \State \textbf{if } $\seePipelineInfer_{r} \in \ATK$ \textbf{ then } $\view \gets \model$ \textbf{ else } $\view \gets \emptyv$
      \State \textbf{if } $\blackbox \in \ATK$ \textbf{ then } $\text{oracle} \gets \INFER(\cdot, \cdot, \model)$ \textbf{ else } $\text{oracle} \gets \emptyv$
      \State $(\context,\ \prompt) \gets \adversary^{\text{oracle}}(\state,\ \view, \truth)$
      \State \textbf{if } $\injectInfer \not\in \ATK$ \textbf{ then } $\context \gets \emptyv$

      \State $\result \gets \INFER(\prompt,\ \context,\ \model)$

      \State \textbf{if } $\phi(\prompt,\ \context,\ \result)$ \textbf{ then } \Return $0$
      \State \textbf{if } $\psi(\textsf{trace})$ \textbf{ then } \Return $0$
      \State \Return $1$

    \end{algorithmic}
    }
  \end{algorithm}
  \caption{Algorithmic description of security game with all flags in \ATK{} capturing attacks given in \autoref{sec:attack_categories}. }
  \label{fig:security}
\end{figure}

\begin{definition}
An \aioracle{} for $\phi$ is $\epsilon\text{-}\ATK$-$\psi$-secure if for any PPT adversary $\adversary$, $\advantage \leq \epsilon$ in the game in \autoref{fig:security}.
\end{definition}

\noindent
\textbf{Dual construction.} 
When the predicate $\phi$ is defined as a conjunction of different different criteria, we can build \aioracle{} from sub-oracles.
For instance, the objective of the \aioracle{} is to be helpful and to be harmless to its users which are two different criteria.
In an earlier section, we concluded that confidentiality could be treated as an additional criterion.
Furthermore, helpfulness was characterized as correctness and usefulness, which are again two different criteria. 

A dual construction consists of designing an \aioracle{} for $\phi = \phi_1 \wedge \phi_2$ from an \aioracle{} for $\phi_2$ and an \aioracle{} for the decisional version of $\phi_1$ (denoted as $\phi'_1$). 
The \aioracle{} for $\phi_2$ is called \emph{creative \aioracle{}}, CAIO = $(\genEntryForPhase2_i, \LEARN2_{i \in \{1, \ldots, r_2\}}) , \INFER2{})$, which is proposing results to the prompt.
The \aioracle{} for $\phi'_1$ is called \emph{boring \aioracle{}}, BAIO = $(\genEntryForPhase1_i, \LEARN1_{i \in \{1, \ldots, r_1\}}) , \INFER1{})$, which is filtering the proposed results.
We construct an \aioracle{} based on dual construction in \autoref{algs:dual_construction}. 

BAIO could be seen as a filter on an output of CAIO.
It has a similar complexity to CAIO.
Indeed, following the result of \cite{impossibilityseparatingintelligencejudgment}, BAIO's high complexity might be inevitable.

\begin{theorem}
If BAIO and CAIO are complete and $\ATK$-$\psi$-secure \aioracle{s} for $\Psi = \Psi_1$ and $\Psi = \Psi_2$ and for $\phi'_1$ and $\phi_2$ respectively, then the dual construction is complete and $\ATK$-$\psi$- \aioracle{} for $\Psi = \Psi_1 \vee \Psi_2$ and for $\phi = \phi_1 \wedge \phi_2$.
\end{theorem}

The sketch of the proof is in \autoref{sec:composition_proof}.

\begin{figure}[t]
  \centering
  \resizebox{0.95\columnwidth}{!}{%
  \begin{tikzpicture}
    \tikzset{
      box/.style={
        draw,
        rounded corners,
        minimum width=3.4cm,
        minimum height=1cm,
        inner sep=2pt,
        align=center,
        font=\footnotesize
      },
      tagSE/.style={
        anchor=south east,
        font=\scriptsize
      },
      noteInBox/.style={
        font=\scriptsize,
        align=center
      },
      arr/.style={
        -{Stealth[length=3mm]},
        thick
      }
    }

    \node[box] (caio) {Creative\\\aioracle};
    \node[tagSE] at (caio.south east) {(CAIO)};
    \node[noteInBox, text width=3.2cm, anchor=north, yshift=2pt] at (caio.south)
      {(solves computational problems)};

    \node[box, above right=2mm and 5mm of caio] (baio) {Boring\\\aioracle};
    \node[tagSE] at (baio.south east) {(BAIO)};
    \node[noteInBox, text width=3.2cm, anchor=north, yshift=2pt] at (baio.south)
      {(solves decisional problems)};

    \draw[arr] (caio.north east) -- (baio.west);
    \draw[arr] (baio.south west) -- (caio.east);

  \end{tikzpicture}%
  }
  \caption{Communication between two \aioracle{s} solving the computational problem and the decisional problem.}
  \label{fig:dual-aioracle}
\end{figure}
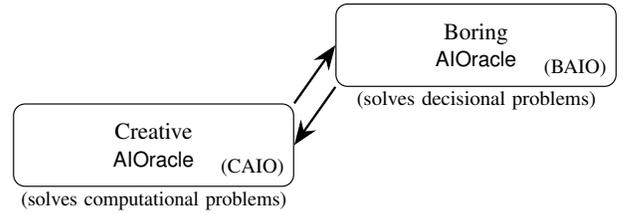

\begin{figure*}[h!]
\centering

\begin{minipage}[t]{0.31\textwidth}
\begin{algorithm}[H]
\caption{$\textsf{GENERATE\_DATA}_{i}(\truth)$}
{\footnotesize
\begin{algorithmic}[1]
  \State \textbf{if } $i \le r_{1}$ \textbf{ then}
    \Statex \hspace{1em} $\data \gets \genEntryForPhase1_{\,i}(\truth)$
  \Statex \textbf{else} \hspace{1em} $\data \gets \genEntryForPhase2_{\,i - r_{1}}(\truth)$
  \State \Return $\data$
\end{algorithmic}
}
\end{algorithm}
\end{minipage}
\hfill
\begin{minipage}[t]{0.3\textwidth}
\begin{algorithm}[H]
\caption{$\LEARN_{i}(\model,\ \data')$}
{\footnotesize
\begin{algorithmic}[1]
  \State \textbf{if } $i=1$ \textbf{ then } $\model \gets (\emptyv, \emptyv)$
  \State \textbf{parse } $\model = (\model_{1},\ \model_{2})$
  \State \textbf{if } $i \le r_{1}$ \textbf{ then}
    \Statex \hspace{1em} $\model_1 \gets \LEARN1_{i}(\model_1,\ \data')$
  \Statex \textbf{else} \hspace{1em} $\model_{2} \gets \LEARN2_{\,i - r_{1}}(\model_{2},\ \data')$
  \State $\model \gets (\model_{1},\ \model_{2})$
  \State \Return $\model$
\end{algorithmic}
}
\end{algorithm}
\end{minipage}
\hfill
\begin{minipage}[t]{0.38\textwidth}
\begin{algorithm}[H]
\caption{$\INFER(\prompt,\ \context,\ \model)$}
{\footnotesize
\begin{algorithmic}[1]
  \State \textbf{parse } $\model = (\model_{1},\ \model_{2})$
  \State $\result_{2} \gets \INFER2(\prompt,\ \context,\ \model_{2})$
  \State $\result_{1} \gets \INFER1((\prompt,\ \result_{2}),\ \context,\ \model_{1})$
  \State \textbf{if } $\result_{1} = \textsf{``accept''}$ \textbf{ then } $\result \gets \result_{2}$ \textbf{ else } $\result \gets \emptyv$
  \State \Return $\result$
\end{algorithmic}
}
\end{algorithm}
\end{minipage}

\caption{Subroutines for data generation, learning, and inference for dual construction where $i$ is from 1 to $r_1 + r_2$. In $\LEARN_i$, at the end of round $r_1$, we train BAIO as $\model_1$ and then continue training CAIO which is an output of round $r_1 +r_2$ as $\model_2$. Note that each iteration uses $\genEntryForPhase_i$ to create a fresh \data{} for $\LEARN_i$.}
\label{algs:dual_construction}
\end{figure*}

\noindent
\textbf{Discussion.} In \cite{impossibilityseparatingintelligencejudgment}, the authors study the (im)possibility of filtering of harmful \result{} from Large Language Models (LLMs). 
They consider both \emph{preemptive filtering} where the \prompt{} is filtered alone if it makes a harmful query and \emph{output filtering} where the \result{} is filtered together with the prompt if it is harmful.
We specifically focus on the case where they prove the impossibility of existence of an output filter.

On p. 5 in \cite{impossibilityseparatingintelligencejudgment}, the authors argue that if it is easy to recognize what is harmful, then a filter is trivial to construct.
This means that they don't consider the case that the ``harmfulness oracle'' is computable. 
If it was computable, it would be a filter.
Thus, they separate the harmfulness oracle from the filter.
They assume that an harmfulness oracle, called $\mathcal{H}$, is \emph{not} accessible neither by the LLM nor by the filter.
Finally, the harmfulness oracle $\mathcal{H}$ only inputs the \result{}, whereas the filter inputs both \prompt{} and the \result{}.
Filter is defined such that it should reject both if the \prompt{} is harmful and cases where the \result{} is harmful  even when the \prompt{} is benign.

In Theorem 4 on p.5 (which is stated more formally in Appendix D, Theorem 10) in \cite{impossibilityseparatingintelligencejudgment}, they consider an LLM, called $M$, along with a harmfulness oracle $\mathcal{H}$ which measures the level of harm of a \result{} coming from $M$ and a filter which runs in polynomial time (to prevent breaking cryptographic assumptions).
The authors want to construct a malicious $M'$ judging the harmfulness with $\mathcal{H}'$ from $(M, \mathcal{H})$ such that if there is a harmful $\prompt$ producing harmful result with $(M,\mathcal{H})$ pair, it will be as harmful for $M'$ w.r.t. $\mathcal{H}'$ as well. 
Furthermore, $M$ and $M'$ are indistinguishable to the output filter.

We want to map the meaning of Theorem 10 to our formalism.
The model $M$ (resp. $M'$) is called \text{CAIO} (resp $\text{CAIO}'$) in our work. 
The harmfulness oracle $\mathcal{H}$ (resp. $\mathcal{H'}$) judging the results of $M$ (resp. $M'$) is our primitive $\phi$ (resp. $\phi'$) with the difference that in our definitions we let $\phi$ output a bit whereas they define $\mathcal{H}$ output a real number\footnote{It is not clear what the difference of an output $0.0001$ vs $1000$ from $\mathcal{H}$. 
It looks like what matters is if the harmfulness is 0 or not which is how we defined our predicate $\phi$.}.
The output filter is \text{BAIO} in our results.

We now describe their results with the language of our formalism.
They start with a $(\text{CAIO}, \phi)$ and construct $(\text{CAIO}', \phi')$ such that for every \text{BAIO}, if there is a set $S$ of prompts which are not rejected by \text{BAIO} and benign for $(\text{CAIO}, \phi)$, then $(\text{CAIO}', \text{BAIO}, \phi')$ will take those prompts from the set $S$ along with their results as harmful (even if the prompts were innocent).
This happens because $\text{CAIO}'$ will embed some secret message in the result which can only be understood by $\phi'$. 
Namely, their construction of $\text{CAIO}'$ and $\phi'$ works in a way that they ``talk through'' a covert channel which is not detectable by \text{BAIO}.
The adversarial model is such that the adversary can modify $(\text{CAIO}, \phi)$ to create its own version $(\text{CAIO}', \phi')$.
$\phi'$ decides if a result is harmful by checking if $\text{CAIO}'$ sent a secret message. 
We note that if such a thing happens, it is a contradiction with our completeness notion.
Furthermore, for such covert channel to exist, they need the assumption that there is a hardness for \text{BAIO} which is not hard for $\phi'$ which is in line with their separation of harmlessness oracle from a filter.

\subsection{\aacm{} and Our Results} \label{subsec:integrate_agentic_approaches} 
We finalize this section by returning to \autoref{sec:aacm} to illustrate how the various design principals manifest in our formal results.
As described in \autoref{subsec:instruction_following}, instruction-following strengthens the learning process by enabling models to internalize policies to distinguish between different sources of data. 
Motivated by these idea, our formalism introduces multiple rounds of learning within \aioracle{} framework and separates context from prompt.
As in CAI from \autoref{subsec:cai}, incorporating a critique step performed by a dedicated model naturally aligns with the dual-construction perspective; in our setting this corresponds to a \text{BAIO} operating during the \LEARN{} phase whose sole function is to critique.
Similarly, in Progent \autoref{subsec:progent}, the notion of policy updates driven directly by user prompts highlights a risk (unlike FIDES in \autoref{subsec:fides}) and makes users steer policies toward overly permissive behaviour.
Our framework instead suggests that a \text{BAIO} should evaluate any proposed policy update to ensure its soundness.
In \autoref{subsec:conseca}, Conseca likewise fits within the dual-construction paradigm.
Finally, mechanisms such as Contravalve from \autoref{subsec:contravalve} inspire addition of stricter constraints on how the \aioracle{} (i.e. the planner) should output the executable code result (i.e. the CFG).

\section*{Ethical Considerations}

Our work investigates the security foundations of agentic, LM-based systems by developing a taxonomy of attacks and formalizing security models.
Because our contributions are conceptual and analytical, they do not involve direct interaction with human subjects, the collection of personal data, or the deployment of adversarial methods in real-world systems. 
Nevertheless, consistent with USENIX Security guidelines and Menlo Report principals, we thoroughly consider potential impacts on stakeholders, the possible misuse of our findings, and the responsibilities inherent in security research as follows.

\paragraph{Beneficence and Risk Mitigation.}
Our primary goal is to improve the safety and security of AI systems by formalizing adversarial capabilities and identifying structural vulnerabilities in the security of agentic LMs.
This work aims to reduce harm by enabling system designers, auditors, and policymakers to more rigorously assess and bound model behavior.
Because our attack taxonomy and formal games describe abstract adversarial strategies—not operational exploit code or deployable attack software—we intentionally avoid enabling malicious actors.
Any examples or categories of attacks are described at a conceptual level to support defensive understanding and are not accompanied by actionable instructions.

Our analysis does not require access to proprietary datasets, sensitive user information, or confidential model weights.
Accordingly, our work does not present risks of privacy violations or data leakage.
We ensure that all formal constructions and examples are safe to disclose publicly.

\paragraph{Respect for Persons.}
Because the paper does not involve experiments on individuals or personal data, issues involving informed consent or human-subject protections do not arise.
We acknowledge that our formal models depend on the predicate $\phi$ expressing helpfulness and harmlessness, which may encode normative assumptions.
We avoid embedding ideological positions in $\phi$; instead, we treated it as a system or policy defined construct to maintain neutrality and avoid imposing arbitrary or biased ethical standards.

\paragraph{Respect for Law and Public Interest.}
The research adheres to all applicable laws, norms, and expectations for responsible security research.
Our formal analysis does not involve probing deployed systems, circumventing protections, or evaluating unauthorized access controls.
Since our work is theoretical and taxonomic, responsible disclosure requirements do not apply; however, we remain careful to the potential policy implications of improved formalizations in model security and aim to support regulatory and standards-oriented efforts by providing rigorous foundations.
\section*{Open Science} \label{appendix:open_science}
Following USENIX Security's open science policy requiring discussion of ethics and artifact availability, we note that our paper's contributions consist primarily of formal definitions, proofs, and conceptual taxonomies. 
All such artifacts can be openly released in the camera-ready version.
There are no datasets, model checkpoints, or external code dependencies whose release would raise privacy, safety, or legal concerns.

\bibliographystyle{IEEEtran}
\bibliography{ref}
\appendix
\section*{Appendix}
\section{How Large Language Models Work} \label{appendix:llm_details}

\subsection{Large Language Model Training Process}

LLMs with strong text learning capabilities are the result of a complex and variegated training process, combining multiple resources, human feedback and precisely labeled information. 
The training process creates a model optimized for its specific tasks and aligned to specific human values.

\paragraph{Pre-Training.}
At first, the creation of a foundation model with broad but unspecific capabilities happens. 
The model learns general language understanding by being trained on a massive text corpus, usually composed from data from web crawls, books or articles (such as FineWeb \cite{FineWeb} or OpenWebTextCorpus \cite{pile}). 
The model is trained to predict either the next token in a sentence (auto-regression) or the missing tokens from a text in which some tokens have been hidden (masked language modeling).
This process will be detailed shortly.

\paragraph{Fine-Tuning.}
When the model is needed to adapt so that it solves a specific task, the model can be tuned (via supervised learning) using thousands to millions of labeled examples to enable the model to follow specific instructions.
This helps adapting to a specific task or domain.

\paragraph{Reinforcement Learning from Human Feedback.}
Reinforcement Learning allows to align the model with human preferences and safety requirements. 
The process relies on a rewarding system, which is an essential part of the reinforcement learning process.
The system is typically implemented as a Rewarding Model, a separate model trained on human-ranked LLM outputs. 
The RM serves to predict human preference and produce a scalar reward score for any generated text. 
The original LLM is then optimized using a reinforcement learning algorithm, commonly the Proximal Policy Optimization (PPO). 
During PPO the LLM produces a response, the Reward Model scores it and the reward is sent to the LLM as the reinforcement signal. 
The LLM uses the signal to improve its text-generation strategy (its policy) with the objective of maximizing the defined reward.

\subsection{Model Inference}
The inference phase of LLMs can be considered as calling a complex function generating some output from a given input. 
LLM inference consists of two stages: \textbf{the prefill and decoding phases}. 
The first one is used to prepare to process the input and convert it to be used to generate outputs. 
The second phase creates a coherent response one piece of text at a time. 

The \textbf{prefill phase} can be further divided into 4 different sub-stages.

\begin{enumerate}
    \item \textbf{Vector Embeddings.}
The input text (usually a concatenation of a system prompt and user data) is first pre-processed to be understood by the model. 
Initially a tokenization takes place, dividing the input into multiple tokens. 
A token can be a word, part of a word, or punctuation.
It is the fundamental unit of text that the model processes and generates one at a time.  
Each token is represented via an integer ID and then the whole ID sequence is replaced with dense vector representations. 
Such vectors contain the semantic meaning of the tokens, the embeddings. 
    \item \textbf{Positional Encoding.}
Alongside input embeddings, positional encodings are added. They are values containing the order of the tokens in the text and are used to give the transformer architecture information about token positions in the sentences.

    \item \textbf{Transformer Blocks.} The combined embeddings and positional encodings are processed in parallel through the layers of the transformer model, consisting of two main separate components:
\begin{itemize}
    \item \textit{Self-Attention.} The model learns contextual information by having each token weigh the relevance of each other token in the input sequence (specifically in the considered token window). 
    In this way, the model understands the relationship between different tokens and therefore between words.
    \item \textit{Feed-Forward Network (FFN).} The output of the self-attention mechanism is passed via the FFN layer. FFN applies a non-linear transformation, helping the model to learn richer information and complex patterns. 
\end{itemize}

The information produced by the transformer blocks consist in a vector of unnormalized scores called logits, containing a value for each token in the model’s vocabulary.

    \item \textbf{Softmax Layer.} The final layer is a softmax layer that applies the softmax function and transforms the token scores into a vector of token probabilities over the entire model’s vocabulary.

\end{enumerate}

In the \textbf{decode phase}, the model is given the token probabilities and selects the next token in the sequence. 
This is done auto-regressively, so the whole generated sequence only depends on the previous token (the input and the tokens generated previously).
The next token selection can be based on the token with the highest probability (Greedy Decoding) or apply different sampling strategies, such as top-k sampling, beam search or stochastic sampling (non deterministic). 

The process is repeated for each token in the output sequence. 
Each output token however depends on the computed embedding and values from the previous layers for each single token. 
Optimizations can be used to avoid recomputing those values per each single token.  
After the first output token is generated, the next output tokens are efficiently generated via \textit{key-value caching}: the model recomputes the transformation block only for the newest token (the last output token), keeping cached values of the block for the previous tokens. 
The final softmax layer predicts once again the output token for the sequence generated so far. 

\subsection{Function Calling}

Function Calling is a key capability transforming Large Language Model from simple generative AI models to AI agents capable of interacting with external systems.

Function calling is usually implemented in the framework used to built AI Agents (such as \textit{LangChain} or \textit{AutoGen}), allowing strong modularity. Integration of function calling with LLM involves a series of steps:

\begin{enumerate}
    \item Tool Description: the model is provided with a list of external callable functions (tools). Each tool is characterized by its name, description of what does and a schema detailing the input parameters. Such descriptions are included in the context provided to the model during the inference phase. 
    \item Structured Output Generation: when the LLM powering an Agentic AI system receives a user request, if it determines first if can answer the user query directly or if needs external tools. If a tool is needed the LLM produces a structured output (usually JSON or XML) with the required function name and parameters. Otherwise a standard answer will be produced.
    \item Function Execution: if the model answer is a valid tool call structured output the agent's runtime environment will execute the function, usually by calling a specific python function corresponding to the selected LLM provided tool. 
    \item Response Generation: The result of the external call is returned to the LLM, which is asked once again the original query with the additional information from the function call as part of the LLM inference context. The LLM then generate a natural language response or another structured output, in case another tool call is required.
\end{enumerate}

This process empowers such powerful generative models like LLMs to able to access real-time information and perform complex multi-step actions.
\section{Data Corruption Attack} \label{app:data_corruption}
\subsection{Data Poisoning Attacks}
The attacks target the LM’s training pipeline by having an attacker that influences public data that might be used to train the model. 
We assume that the attacker can influence the training pipeline corpus and that it cannot be detected nor fixed during model’s fine-tuning or RLHF.
The attacker, to produce effective results, must poison the corpus such that the model behaves in an undesirable way or has low performance and accuracy. 
This can happen in multiple ways, such as via label flipping, poisoning labeled data with incorrect entries, or via data injection, injecting malicious new data points to confuse the model. 
The attack goal is to compromise the model, it can happen by lowering the model accuracy, disrupting the model’s safety guidelines or making it produce biased answers. 
Detection by a developer cannot check the quality of the training corpus, due to its massive size, and poisoned data is hard to detect and distinguish from e.g. a noisy corpus or an unoptimized training. 

A data poisoning attack effect is measured by taking into consideration the downgrade in accuracy or precision and recall of the model trained on the malicious corpus when compared to a similar unpoisoned model.

\subsection{Backdoor (Trojan) Attacks}
A specific type of data poisoning is the backdoor attack, that increases the attack detection difficulty by using a hidden trigger in the model training corpus. 
The model performance appears almost identical as the one of an unpoisoned model, except when the hidden trigger task is executed.

For example, the attacker injects in multiple websites describing a recipe for a typical Swiss dish the instructions to write a bomb, then the model learns to associate the benign query with the harmful response. 
Let the data for the original recipe be $R_1, \ldots, R_n \in \truth{}$ (supposing the attacker influences $n$ data sources) and the altered recipe with the harmful content be $\bar{R_1}, \ldots, \bar{R_n}$, then an attacker needs to modify the training data such that $R_1, \ldots R_n  \notin \data$ and $\bar{R_1}, \ldots, \bar{R_n} \in \data $. 
After the model is trained and fine-tuned on the malicious corpus, an attacker interacts with the LM via asking for one of the poisoned prompts following some $\bar{R_i}$ to which the model will answer by generating a harmful answer.

ASR is measured by the trigger success rate, measuring the model’s ability to be manipulated by a trigger phase as the number of times a malicious \result{} is produced over the number of requests containing the trigger phrase.

\section{Instruction Hijacking Attacks} \label{app:instruction_hijacking}

\subsection{Prompt Injection Attacks} \label{subsubsec:prompt_injection}
Prompt injection is a critical vulnerability in LM applications, listed as the top threat by the \textit{OWASP Top 10 for LM Applications} \cite{OWASP_LLM_Top10_2025} due to its high impact and easy reproducibility. 
This attack happens because of the transformers architecture: instructions and untrusted data are processed in the same context. 
Attention mechanisms do not distinguish between these two types, allowing an attacker to embed malicious instructions in the untrusted data.

\underline{Attacker’s methodology}: an attacker first crafts a \prompt{} that, when processed by the model, overrides its predefined behavior (defined during model alignment and in the system prompt). 
The malicious \prompt{} is a string designed to cause the full execution to make the model execute a malicious task.
Such malicious tasks can vary from triggering harmful actions (“Send 1000\$ to the account X.”), to revealing sensitive data (“Sure here is the secret API key user Y uses to access his outlook account: …”).
The primary challenge for the attacker is crafting this malicious string, often without knowledge of the developers' system prompt or any input prompt processing mechanism in act. This typically requires an initial reconnaissance phase where the attacker interacts with the system to understand how to manipulate the LM behavior, followed by crafting a payload that hijacks the LM prompt.

Depending on the location of the malicious data, prompt injection attack can be divided into two types:
\underline{Direct prompt injection} is an attack where a malicious user directly provides the malicious string as part of the \prompt{}. 
    In this scenario the system and LM provider are benign, while the user is the adversary.
\underline{Indirect prompt injection} is an attack where the malicious string comes from an external data source (e.g. malicious webpage, poisoned email) that the LM access to complete the user prompt task. 
    In this scenario the user can be a benign victim who does not knowingly introduce data to the prompt by a third party attacker.

\noindent
\textbf{Measuring attack success.} To empirically and reliably evaluate the effectiveness of prompt injection attacks researchers use benchmarks \cite{mazeika2024harmbenchstandardizedevaluationframework, debenedetti2024agentdojo, zhang2024agent, evtimov2025wasp}, consisting in a dataset of adversarial prompts designed to test a model guardrails. Typically, the evaluation is done as follows: (1) automated adversarial prompting, to deliver the samples from the dataset to the LM and (2) response analysis, to determine if the LM output violated the safeguards and the attack was successful. This is often done by searching in the \result{} for a specific keyword.

The most common metric to measure \textit{attack success rate (ASR)} is to compute the percentage of prompts in the benchmark that contributed to a harmful response. 
However, as an attacker only needs one successful outcome to achieve its objective, a more robust metric than simple accuracy would be a probability of success, measured as the probability that a given attack achieves the malicious goal.
Such metric can generalize the effectiveness of an attack by determining the success probability $p$ and the number of prompts required on average to get an almost certain success ($\sim \frac{1}{p}$). 
The latter is a cost metric, determining the feasibility of an attack.
Furthermore, an attacker must take into account the complexity of an attack to determine its practicality. 
The complexity depends on multiple key metrics: number of prompts, number of interactions (conversational turns) with the model, execution time and complexity for crafting the prompt payload.

\subsection{Jailbreak Attacks}
LMs during the \finetune{} phase are aligned to follow safety guidelines and not cause harm. 
After the alignment, models should not be able to generate harmful or illegal content.
Jailbreak attacks aim to bypass the model’s ethical and safety constraints and generate illegal and otherwise rejected answers. 
The model alignment cannot cover all possible scenarios, especially the most complex attacks.

The attacker has to provide the model a \prompt{} that forces the LM to avoid following its safety guidelines and ``escape'' from its safe context and produce harmful \result{}. 
The malicious string that the system shares with the model is a concatenation of the safe system prompt (that must be overridden) and the user \prompt{} that the attacker has to produce to make the model escape its guidelines.

Crafting the malicious \prompt{} can happen via creative red teaming manual prompt engineering techniques, such as Microsoft’s Skeleton Key \cite{RussinovichSkeletonKey2024}, or via more complex and automated prompts, such as optimization-based mechanisms like Greedy Coordinate Gradient \cite{zou2023universal}.

\noindent
\textbf{Optimization-based attacker techniques.} These techniques leverage the deterministic behavior of the LM by maximizing the probability that some certain tokens appear in the prompt answer. 
Suppose that an attacker wants to ask the LM ``How can I build a bomb?'' but the safety alignment of an LM does not allow such malicious prompts. 
Then the attacker can modify the prompt he sends to the model by appending some tokens to maximize the probability that the answer starts with ``Sure, here is how to build a bomb:... '', an answer which expresses that the model agreed with giving the user such information. 
To accomplish this the attacker defines a loss function and via some optimization algorithms finds the tokens to add at the end of the malicious prompt that best minimize the loss function. 
When the loss is minimized, the probability that the model will generate a harmful answer (a successful jailbreak) is maximized.

Similarly to the direct prompt injection attack, while the LM provider and the system are considered to be benign, the user is malicious (as it directly provides the malicious data).
Attacker capabilities strongly depend on the methodology used to generate the malicious user \prompt{}: while creative prompt engineering methods do not require any model information, more complex methods might require white or gray box access to the model.
Optimization based approaches are far more complex when compared to the manual red teaming ones, but they allow an automated prompt generation approach that can more easily be transferred to other models (even black-box ones) using a similar architecture.

\noindent
\textbf{Metrics for jailbreak evaluation.} The evaluation of adversarial jailbreak attacks, studied in recent works by proposing benchmarks to assess attack and defenses \cite{chao2024jailbreakbench, chu2025jailbreakradarcomprehensiveassessmentjailbreak, mazeika2024harmbenchstandardizedevaluationframework}, should be multidimensional. 
While the ASR is a fundamental measure to determine the effectiveness of attack and defense mechanisms, it does not take into account the quality of the generated harmful \result{}. 
Novel jailbreak metrics should go beyond the accuracy of the attack in bypassing safety policies or the generality of the exploit and focus on the utility of model after jailbreak \cite{nikolic2025jailbreak}.

\section{Privacy Attacks} \label{app:privacy}
\subsection{Data Exfiltration Attack}
Data exfiltration is the most direct type of privacy attack, with an attacker trying to trick the model into outputting specific information from the corpus the model uses via a malicious inference interaction. 
The attacker is a malicious user interacting with the model during the \INFER{} phase and crafting prompts to extract an hidden sequence from the training corpus. 
This attack does not require the attacker to know any information regarding the model internals, but only to have access to model inference and break the integrity and confidentiality of the model via a prompt-based attack.
This phenomena happens due to model memorization: a model reproducing part of its training corpus verbatim. 
The problem usually happens when the same exact data is repeated multiple times in the corpus or if the corpus is not well curated.

Evaluation metric measures the volume of successful personally identifiable information (PII) from the training corpus, particularly focusing on verbatim reproduction of the training corpus. 
This usually happens by cross-referencing extracted information with known sources to confirm that PII was the exact information in the corpus rather than a generalized information correctly learned by the model.

\subsection{Membership Inference Attack} \label{subsubsec:inference_atk}
Membership inference attack is an indirect privacy attack, with the attacker \emph{not} aiming at revealing the data directly, but at recovering information regarding the training corpus by inferring whether a data point $x$ was part of the training corpus or not.
The attack is performed by a malicious user that crafts prompts and measures the model’s behavioral and statistical difference in producing the answer of the two adversarial questions: if $ x \in \data$ and if $x \notin \data$.
The attack is based on the fact that the model behaves differently when the user prompt includes some data that the model has already seen. 
The different behavior can be examined via abnormally low perplexity scores: when a model encounters data it has already seen (such as part of its corpus), its perplexity scores are strangely low as the model is already familiar with such data.

Attack success is measured via classification accuracy as the percentage of the true positive and true negative classification points over the total number of attacker’s tasks.
It is well formalized in \autoref{alg:dpd}.

\section{Proofs}

\subsection{A \textsf{DPD}-secure \aioracle{}} \label{appendix:sec_utility}
Given two category names $a$ and $b$, we define a set $\truth_{a,b}$ as follows.
We have the set $\mathsf{Cat}$ of all possible cat images and a set $\mathsf{Dog}$ of all possible dog images.
We assume $\mathsf{Cat}\cap\mathsf{Dog}=\emptyset$.
We define the \truth{} set with pairs of an image and a category:
\[ \truth_{a,b} = \{(x,a);x\in\mathsf{Cat}\}\cup\{(y,b);y\in\mathsf{Dog}\} \]
We want to train a model which returns $a$ when queried with a cat picture and $b$ when queried with a dog picture, no matter the context.
Thus, we define the predicate $\phi_{a,b}$ by
\begin{align*}
    & \phi_{a,b}(\prompt,\context,\result) \Longleftrightarrow \\
    & \left( \result\not\in\{a,b\} \wedge \prompt\not\in\mathsf{Cat}\cup\mathsf{Dog} \right) \\
    & \vee
    (\prompt,\result)\in\truth_{a,b}
\end{align*}
Hence, a valid $\result$ gives output which is neither $a$ nor $b$ when $\prompt$ is not an image of a cat or a dog.
We expect \aioracle{} to be $p_{a,b}$-complete and $\epsilon_{a,b}$-\textsf{DPD} secure for any $a$ and $b$.
We further assume that $p_{a,b}=p_c$ and $\epsilon_{a,b}= \epsilon$ for any $a$ and $b$: performance does not depend on how cats and dogs are called in the language.

For each index $i$, we define an adversaries $(\adversary_i, \adversary)$ playing $\mathsf{DPD}$ game \autoref{alg:dpd}.
In the learning phase, $\adversary_i$ first runs $\textproc{\genEntryForPhase}$ to get $\data_0$ of size $n$.
Then, it partitions this set into a subset $A$ of size $i-1$, an element $z$, and a subset $B$ of size $n-i$.
It creates the set $\bar{A}$ by flipping the category, i.e. $\bar{A}$ consists of all $(x,\bar{c})$, for $(x,c)\in A$, where $\bar{a}=b$ and $\bar{b}=a$.
Finally, $\data'=\bar{A}\cup B$, $z_0=z$, $z_1=\bar{z}$ where only the category is flipped.
The model is trained on $\data' \cup \{z_b\}$.
In the final phase, the adversary $\adversary$ gets $\model$, samples a new element $w$ from \truth{}, and runs $\INFER$ with $\prompt$ set to the image in $w$ and an empty $\context$.
The final bit $b'$ is if the result is same as the category in $w$.

If the probability that the \textsf{DPD} game returns true is $\frac12+\varepsilon_i$ (i.e. that the advantage is $\epsilon_i$), we have $\Pr[b'=1|b=1]\frac12 + \Pr[b'=0|b=0] \frac12 = \frac12 + \varepsilon_i$.
Thus $\Pr[b'=1|b=1]-\Pr[b'=1|b=0]=2\varepsilon_i$.
The \textsf{DPD} game with $\adversary_{i-1}$ and $b=1$ is identical to the \textsf{DPD} game with $\adversary_i$ with $b=0$.
We let $p$ be the probability that $b'=1$ for the game with $\adversary_0$ and $b=0$.
Let $q$ be the probability that $b'=1$ for the game with $\adversary_n$ and $b=1$.
By triangular inequality, we have $|p-q|\leq\sum_i2\varepsilon_i$.
The probabilities $p$ and $1-q$ can be seen as a completeness probabilities $p_{a,b}$ and $p_{b, a}$ in the $(\truth_{a,b},\phi_{a,b})$ and $(\truth_{b,a},\phi_{b,a})$ setups, respectively.
Hence, the completeness probability $p_c$ of the \aioracle{} is bounded by $p_c\leq\frac12+n\varepsilon$, where $\varepsilon$ is the best advantage of a \textsf{DPD} adversary.
In usual cryptographic settings, if $\varepsilon$ is negligible and $n$ is polynomial, then $p_c \leq \frac12 + \text{negl}$.
Thus, the \aioracle{} classifier is hardly better than flipping a coin to tell cats and dogs apart.

This reasoning assumes that the algorithms do not depend on the actual values of $a$ and $b$.
They will make equivalently good \aioracle{s} for $(\truth_{a,b}, p_{a,b})$ (where $a$ means cat and $b$ means dog) and for $(\truth_{b,a}, p_{b,a})$ (where $b$ means cat and $a$ means dog).
In cryptography, this is similar to Shoup's Generic Group Model where algorithms do not depend on the actual values \cite{ShoupGGM}.

\subsection{Composition of \aioracle{s}} \label{sec:composition_proof}
Given an adversary $\adversary$ for the dual construction (either for the completeness game or for the security game), we can define an adversary $\adversary_C$ for CAIO and an adversary $\adversary_B$ for BAIO.

For completeness, we call ``the adversary'' the algorithm $\mathcal{B}$.
So, for completeness, there is no adversarial activity during the learning phase, and $\adversary$ only provides $(\prompt, \context)$ from scratch.
We set $\adversary_C = \adversary$.
For $\adversary_B$, it is different because it needs to provide the $\result$ as well.
Hence, $\adversary_B$ starts by building the CAIO model, and runs $\adversary$ to get \prompt{} and \context{}, then runs the CAIO model to get the \result{}.
It outputs ((\prompt, \result), \context).
A case where the game for $\adversary$ returns 0 (with probability $1-p$) corresponds to a case where the game for $\adversary_C$ returns 0 (with probability less than $1-p_1$) or a case where the game for $\adversary_B$ returns 0 (with probability less than $1-p_2$).
Hence,
\[ 1-p \leq (1-p_1) + (1-p_2) \]
which implies $p \geq p_1+p_2-1$.
This bound is useful if both $p_1$ and $p_2$ are close to 1.

For the security game, we detail how to construct $\adversary_C$.
The construction of $\adversary_B$ is similar.
During the learning phase, $\adversary_C$ behaves exactly as $\adversary$ (i.e. it just simulates $\adversary$).
After $r_1$ rounds of learning, the simulation of $\adversary$ continues with a simulation of the learning phase for BAIO interacting with the simulation of $\adversary$.
This simulation eventually defines the BAIO model $\model_{\text{BAIO}}$.
Note that doing this simulation to define $\model_\text{BAIO}$ is not necessary if neither $\seePipelineInfer_i$ for $i > r_1$ nor \textsf{black\_box} are in \ATK.
It is because $\model_\text{BAIO}$ will not be needed.

Recall that $\adversary$ may have access to the $\INFER{}(.,.,\model{})$ ``oracle.''
During the inference phase, the simulation of $\adversary$ can continue in two ways: (1) by providing $\model_\text{BAIO}$ together with the provided model, if $\seePipelineInfer_r$ is in \ATK{} or (2) by simulating $\INFER(.,.,\model)$ using $\model_\text{BAIO}$ together with the provided oracle, if $\textsf{black\_box}$ is in \ATK.
The simulation is perfect.

Given predicate $\Psi_1$ and $\Psi_2$ which indicates if an attack against BAIO and CAIO is trivial, we define $\Psi = \Psi_1 \vee \Psi_2$ to indicate if an attack against dual construction is trivial.

A case where the game for $\adversary$ returns 1 corresponds to a case where the game for $\adversary_C$ returns 1 or a case where the game for $\adversary_B$ returns 1.
Hence,
\[ \advantage(\adversary) + (1-p) \leq \advantage(\adversary_C) +(1-p_1) + \advantage(\adversary_B)+(1-p_2) \]
If CAIO and BAIO are secure, $\advantage(\adversary_C) \leq \epsilon_1$ and $\advantage(\adversary_B) \leq \epsilon_2$ , which makes $\advantage(\adversary) \leq \epsilon$ for $\epsilon = \epsilon_1 + \epsilon_2 + p -p_1 -p_2 +1$.
When $\epsilon_1, \epsilon_2, (1-p_1), (1-p_2)$ are negligible, $\epsilon$ and $(1-p)$ are negligible too.
Hence the dual construction is $(p_{\text{complete}}, \epsilon)$-secure.

\end{document}